\documentclass[twocolumn,tighten]{aastex63}

\received{2021 January 8}
\revised{2021 April 19}
\accepted{2021 May 18}
\published{2021 July 16}
\submitjournal{ApJ}

\shorttitle{Pulsar Radio Emission by Solitons}
\shortauthors{Ben\'a\v{c}ek et al.}

\usepackage{amsmath}

\begin{document}

\title{Radio Emission by Soliton Formation in Relativistically Hot Streaming Pulsar Pair Plasmas
}

\correspondingauthor{Jan Ben\'a\v{c}ek}
\email{benacek@tu-berlin.de}

\author[0000-0002-4319-8083]{Jan Ben\'a\v{c}ek}
\affiliation{Center for Astronomy and Astrophysics, Technical University of Berlin, 10623 Berlin,
Germany}

\author[0000-0002-3678-8173]{Patricio~A.~Mu\~noz}
\affiliation{Center for Astronomy and Astrophysics, Technical University of Berlin, 10623 Berlin,
Germany}

\author[0000-0002-6900-5729]{Alina~C.~Manthei}
\affiliation{Physikalisches Institut, Rheinische Friedrich-Wilhelms-Universit\"at Bonn, 53115 Bonn, Germany}

\author[0000-0002-5700-987X]{J\"org~B\"uchner}
\affiliation{Center for Astronomy and Astrophysics, Technical University of Berlin, 10623 Berlin,
Germany}
\affiliation{Max Planck Institute for Solar System Research, 37077 G\"ottingen, Germany}



\begin{abstract}
A number of possible pulsar radio emission mechanisms are based on
streaming instabilities in relativistically hot
electron-positron pair plasmas. At saturation the unstable waves
can form, in principle, stable solitary waves which could emit the
observed intense radio signals.
We searched for the proper plasma parameters which would
lead to the formation of solitons,  investigated their properties
and dynamics as well as the resulting oscillations of electrons and
positrons possibly leading to radio wave emission.
We utilized a one-dimensional version of the relativistic
Particle-in-Cell code ACRONYM initialized with an appropriately parameterized
one-dimensional Maxwell-J\"uttner velocity space particle distribution
to study the evolution of the resulting streaming instability in
a pulsar pair plasma.
We found that strong electrostatic superluminal L-mode solitons are formed
for plasmas with normalized inverse temperatures $\rho \geq 1.66$
or relative beam drift speeds with Lorentz factors $\gamma > 40$.
The parameters of the solitons fulfill the wave emission conditions.
For appropriate pulsar parameters the resulting energy densities
of superluminal solitons can reach up to $1.1 \times 10^5$~erg$\cdot$cm$^{-3}$,
while those of subluminal solitons reach only up 
to $1.2 \times 10^4$~erg$\cdot$cm$^{-3}$.
Estimated energy densities of up to $7 \times 10^{12}$~erg$\cdot$cm$^{-3}$
suffice to explain pulsar nanoshots.
\end{abstract}

\keywords{Radio pulsars --- Plasma physics --- Magnetic fields --- Computational methods}


\section{Introduction} \label{sec:intro}
Pulsars are neutron stars that are observed as very bright objects at radio wavelengths.
Their emission is thought to be generated in the strongly-magnetized,
relativistically hot electron-positron pair plasmas of their
magnetospheres.
Pulsars have already been studied for more than fifty
years~\citep{Kramer2002,Melrose2017b,Beskin2018}.
A number of models has been conjectured to
explain the formation of their radio
signals~\citep{Sturrock1971,Ruderman1975,Usov1987,Petrova2009,Philippov2020,Melrose2020b}.
Nevertheless, there is still no consensus on a  proper
emission process that explains all the observed
radio wave features~\citep{Melrose2017a,Melrose2020a}.

The current standard emission models are based on the creation,  
transport and instabilities of a relativistic pair plasma in the
pulsar magnetosphere~\citep{Cheng1977b}.

The pulsar rotation period ranges from milliseconds up to a few seconds.
Since the plasma of the pulsar magnetosphere is ideally conducting 
(negligible resistivity), magnetic fields are frozen into the plasma 
so that $\boldsymbol{E} \cdot \boldsymbol{B} = 0$ 
holds.
On open field lines, on the other hand, 
finite parallel electric fields  $\boldsymbol{E}_\parallel \neq 0$, 
(i.e. $\boldsymbol{E} \cdot \boldsymbol{B} \neq  0$)
may exist, forming so called ``gaps''.  
The number density of charged particles, which have to 
compensate the electric field caused by the fast rotation 
of the magnetospheric plasma, is the so called 
``Goldreich-Julian density''~\citep{Goldreich1969}.
The electric field component parallel to the open magnetic 
field lines above the pulsars polar cap is not compensated 
by the electric currents due to the Goldreich-Julian charging.
They can accelerate particles which can escape the pulsar
magnetosphere.

These accelerated particles form a (``primary'') beam with a typical relativistic Lorentz
factor $\gamma \sim 10^6 - 10^7$.
Due to their motion along curved field lines 
the beam particles emit $\gamma$-ray photons.
In the strong magnetic field, assuming field strength
of the order  $10^{12}$~G~\citep{Arendt2002}
these photons create electron-positron pairs,
which form a
``secondary'' beam with typical Lorentz factors of
$\gamma \sim 32 - 178$ in the pulsar reference
frame .
According to 
\citet{Eilek2016,Mitra2017,Melrose2020a,Philippov2020,Rahaman2020},
the observed pulsar radio signals are assumed to be generated
by these secondary electron-positron particle beams. 
\par
Three main groups of possible mechanisms have been 
suggested to explain pulsar radio emissions due to
those secondary beams:

\begin{enumerate}
\item
One group of emission mechanisms is related
to (either coherent or incoherent) curvature
radiation~\citep{Ruderman1975,Buschauer1977,Melikidze2000,Gil2004,Mitra2017}.
Curvature radiation can be emitted in vacuo and therefore it is
independent from the plasma properties.
In the strong pulsar magnetic field the plasma is characterized by a vanishing perpendicular momentum, i.e., $p_\perp = 0$.
The radiation due to curved magnetic field lines is confined
into a narrow cone. Coherent curvature emission
requires a coherent emission of whole bunches of
charged particles.
Incoherent curvature radiation, on the other hand, is unlikely
to cause the observed coherent radio pulses~\citep{Beskin1993}
as the emitted flux cannot explain such observations.
Indeed, incoherent curvature radio emission would cause
brightness temperatures lower than $10^{13}$~K \citep{Melrose1978}.
Hence, this mechanism cannot explain the observed brightness
temperatures of up to $10^{41}$~K \citep{Soglasnov2004,Jessner2005,Hankins2007}.
Although coherent curvature emission as cause of the fine-structure radio pulses has been discarded due to many reasons in several publications ~\citep{Melrose1981,Lesch1998,Melrose2017a},
it still remains as one of the proposed mechanisms for pulsar radio emission.

\item 
The second group of conjectured emission mechanisms is
related to the
relative drift of plasma particles, which causes relativistic
streaming (beam) instabilities. These instabilities lead to the generation of
plasma waves, which, in turn, can be transformed into electromagnetic waves
escaping from pulsar magnetospheres~\citep{Melrose1999}.
Instabilities of three possible plasma streaming scenarios have been considered by the models of this group:

\begin{enumerate}
    \item The relative streaming between the primary and secondary beams.
    The resulting instability has, however, been found to be inefficient
    in the lower pulsar magnetosphere where it is supposed to occur,
    since its growth rate is estimated to be too small ~\citep{Cheng1977a,Buschauer1977,Arons1981}.
    
    According to \citet{Usov1987}, the development of this streaming instability takes $\tau_\mathrm{I} \sim 10^{-4} (r/R)^{\frac{3}{2}}$~s
    (here $r$ is the height where the emission occurs and $R$ is the neutron star radius).
    By contrast, particles travel this distance in a much shorter time of
    $\tau_0 \sim 3 \times 10^{-5} (r/R)$~s $ < \tau_\mathrm{I}$.
    Hence, this instability cannot evolve, which can be traced back to the too large Lorentz factor and too small density of the primary beam.

    \item The acceleration of the secondary-beam electrons and positrons due to the net parallel current density along the curved magnetic field lines causes a differential streaming of these two populations~\citep{Cheng1977b,Weatherall1994}.
    
    Due to the relativistic temperature of the magnetospheric plasma, which leads to a large thermal spread in the electron and positron velocity distributions, this kind of streaming instability is also not likely to produce the observed radiation~\citep{Buschauer1977}.

    \item The relative streaming between bunches of the secondary-beam electrons and positrons.
    Such particle bunches are thought to be formed in the polar-cap gap regions
    during the so-called ``spark'' events, which create the secondary-beam populations via pair production ~\citep{Ruderman1975,Cheng1977b}.
	The secondary-beam particle density increases during 
    those events, eventually screening out the parallel electric field in the gap  regions and consequently reducing particle acceleration and therefore the pair production frequency.
	This process continues until the secondary beam density 
	reaches the Goldreich-Julian density and particles are released from this region.
	It becomes again almost empty and consequently the parallel electric fields are re-established
    The typical particle-release and electric-field-reformation times are
    up to a few microseconds.
    The particles in each bunch have a large spread in velocity space. 
    In the interaction of bunches, the fastest particles of the
    following (trailing) bunch eventually catch up the slower particles of
    the preceding bunch~\citep{Usov1987,Ursov1988,Usov2002}.
    This causes a two-stream instability.
    The interaction occurs at a typical height of $10^8$~cm above
    the pulsar surface, which agrees 
    in order of magnitude with the results of 
    \citet{Mitra2017}, who concluded that the radio emission
    originates from $200-500$~km above the star.
    \citet[Appendix A]{Melrose2020a} discussed the possibilities of 
    overlapping (overtaking) of particles in the phase velocity space.
    They concluded that such overlapping is not possible due to 
    the fact that there is a spatial elongation of plasma distribution 
    and the distance needed for an overlap of the bunch particles 
    would exceed the typical pulsar emission height.
    Moreover, there are difficulties with analytical predictions of 
    the bunch interaction, the density ratio of the beam and 
    background, the kinetic energy separation of beam particles 
    and background plasma for the calculation of the relativistic plasma 
    dispersion properties for such distributions.
\end{enumerate}
\citet{Rahaman2020} found that the growth rates of streaming 
instabilities for specific distributions are larger for 
scenario (b) than for (c).
The growth rates, they calculated for scenarios (b) and (c) 
are, however, significantly smaller than those obtained 
by~\citet{Manthei2021}, who re-analyzed scenario (c) in detail.

\item
The third group of conjectured pulsar radio wave emission
considers the oscillations
of particles that attempt to screen out the parallel component of the induced electric
field~\citep{Beloborov2008,Lyubarsky2009,Timokhin2013}.
In fact, \citet{Melrose2020b} recently found that oscillations can indeed
produce longitudinal superluminal waves able to escape
from the magnetosphere as O-mode waves.
\end{enumerate}

Regarding the second group of mechanisms mentioned above
(particle streaming causing plasma instabilities), three possibilities exist for the conversion of the generated waves into
electromagnetic waves that could cause the observed pulsar
radio signals~\citep{Melrose1995,Eilek2016,Melrose2017b}:
Relativistic plasma emission, linear acceleration emission
(free electron maser), and anomalous Doppler emission:

\begin{enumerate}
\item  Relativistic plasma emission is due to modulational instabilities that
transform longitudinal electrostatic plasma waves propagating in the
direction of the ambient magnetic field into escaping electromagnetic
waves~\citep{Weatherall1997,Weatherall1998}.
In this case the force acting on the particles is perpendicular to the particle motion.

\item In the  linear acceleration (free electron maser) emission
mechanism, the force acting on the particles
is due to the electric field component parallel to the direction of particle motion.

\item[] Both mechanisms assume the same particle oscillation
frequency as the wave frequency~$\omega_0$ and emission at higher frequencies
$\gamma^2 \omega_0$ \citep{Cocke1973,Melrose1978,Kroll1979,Fung2004,Levinson2005,Melrose2009a,Melrose2009b,Reville2010,Timokhin2013}.
For simplification it is usually assumed that the plasma frequency is the same as the oscillation frequency
(e.g. \citet{Eilek2016}).
The origin of such high frequency oscillations is, however,
still unclear.
As the Lorentz factor of a particle changes significantly during
one oscillation, this mechanism seems to be more relevant
for emission at frequencies much higher than the plasma
frequency.

\item The anomalous Doppler emission applies for particles
which fulfill the resonance condition $\omega-kv=n\omega_\mathrm{ce}/\gamma$,
where $n=-1$ is the harmonic number of the resonance, 
$\omega_\mathrm{ce}$ is the electron cyclotron frequency,  $v$ is the particle velocity, 
$\omega$ is the wave frequency, and $k$ is the wavenumber.
\citep{Machabeli1979,Kazbegi1991,Lyutikov1999a,Lyutikov1999b}.
The emission frequency is
$\omega \approx 2 (v_\mathrm{A}/c)^2 \omega_\mathrm{ce} / \gamma$,
where $v_\mathrm{A}$ is the Alfv\'{e}n speed and $c$ is the speed of light.
These frequencies are, however, only for small values of $\omega_\mathrm{ce} / \gamma$ in the radio band \citep{Melrose2017b}.
\end{enumerate}

Note that all these emission mechanisms are closely related to the emission
of electromagnetic waves by charged particles moving either in a turbulent 
plasma or in structures like solitary waves (solitons).
\citet{Hinata1976}, e.g., proposed that a electron-positron pair-turbulence
could cause 
coherent electromagnetic waves.
\citet{Medvedev2000} developed a ``jitter'' radiation model of ultrarelativistic
electrons moving in nonuniform small-scale magnetic field structures.
Such a radiation could take place when the angle associated 
with
the transverse electron motion in those magnetic fields is smaller than the beaming emission angle.
It is different from the synchrotron radiation, which takes place if the opposite 
inequality between those angles holds.
The theory of particle radiation was extended to media with random
inhomogeneities by \citet{Fleishman2006,Fleishman2007} while
\citet{Workman2008} discussed the difference between synchrotron
and jitter radiation.
\citet{Teraki2014} studied the radiation spectra caused by relativistic
electrons moving in Langmuir turbulence.
\citet{Melikidze2014} analyzed how the electron motion in solitary
waves can cause electromagnetic emission.
The development of strong plasma turbulence and the formation of
solitary wave packets was suggested to explain pulsar nanoshots and
microbursts~\citep{Eilek2016}.
\citet{Weatherall1997,Weatherall1998} described electromagnetic
emission due to a modulational instability of Langmuir waves.
The generation of solitary waves in pulsar magnetospheres was
analytically studied by \citet{Weiland1978,Fla1989,Misra2015,Misra2017,Chaudhuri2018,Carlevaro2020}.
\citet{Melikidze2000} suggested that coherent pulsar radiation
could be emitted from solitary waves with a typical size of
$10-100$~cm ($\sim1.4 - 14\, d_\mathrm{e}, d_\mathrm{e} = c / \omega_\mathrm{p}, \omega_\mathrm{p} = 4.2 \times 10^{9}$~s$^{-1}$).
Each soliton would be associated with an electric charge of
$Q = e \cdot 10^{21}$.
Up to $10^{5}$ solitons could be released every $10\,\mu$s
in 10 independent sparking regions.
This process could generate the observed radio power
of $10^{29}$~erg$\cdot$s$^{-1}$.
While \citet{Melikidze2000} described the evolution of
the solitary waves by solving a nonlinear Schr\"odinger
equation, ignoring Landau damping, \citet{Lakoba2018}
analyzed the consequences of Landau damping confirming
that this effect also allows for the formation of solitary structures.
For the considered parameters, \citet{Melrose1999,Melrose2020a}
found that the growth rate of the streaming instability of a
relativistic pair plasma would be too small for an efficient
pulsar radio emission.
\citet{Manthei2021}, however, showed that for some pulsar plasma parameters
the streaming instability can develop fast enough
during the sparking time periods of 10~$\mu$s.
In particular, the plasma temperature appears to be decisive
for allowing a sufficiently fast growth of the streaming
instability. 
In fact, it has been shown by \citet{Arendt2002}
that the pulsar pair plasmas are so hot that their velocity
space distribution functions should obey a relativistically
covariant Maxwell-J\"uttner function~\citep{Juttner1911}
for both beam (subscript 1) and background
(subscript 0) plasma~\citep{Rafat2019b}:
\begin{equation} \label{eq1}
 g(u) = g_0(u) + g_1(u) ,
\end{equation}
\begin{equation} \label{eq2}
 g_0(u) = \frac{n_0}{2 K_1(\rho_0)} \mathrm{e}^{- \rho_0 \gamma},
\end{equation}
\begin{equation} \label{eq3}
 g_1(u) = \frac{1}{\gamma_\mathrm{b}}\frac{n_1}{2 K_1(\rho_1)} \mathrm{e}^{- \rho_1 \gamma_\mathrm{b} \gamma (1 - \beta \beta_\mathrm{b})},
\end{equation}
where $n_0, n_1$  are the number densities of the background and beam
plasma, respectively, $u/c = \gamma \beta$ and $u_\mathrm{b}/c = \gamma_\mathrm{b} \beta_\mathrm{b}$
are the plasma four-velocity and the mean beam four-velocity,
respectively, normalized to the speed of light $c$, with
$\gamma = (1 - \beta^2)^{-\frac{1}{2}}$ and $\gamma_\mathrm{b} = (1 - \beta_\mathrm{b}^2)^{-\frac{1}{2}}$
their Lorentz factors.
$\rho_0$ and $\rho_1$ are the inverse background and
beam temperatures in units of the particle rest energy
$\rho_i = mc^2 / T_i, \, i=0,1$, $m$ is the electron mass and $T_i$ is the
temperature of the $i$-th species.
$K_1$ is the MacDonald function of first order (modified Bessel function of
the second kind).
Within the~\citet{Usov1987} model, the background bulk velocity is
$\langle u \rangle = 0 $ in the reference frame of the slow particles from the first bunch,
while the fast particles of the following bunch form a beam with $\langle u \rangle \neq 0 $.
The distribution is effectively one-dimensional in the direction of the
pulsar magnetic field. In fact, the perpendicular velocity does not
play any significant role since any perpendicular particle motion
immediately emits synchrotron radiation and vanishes at time scales
much smaller than the plasma period~\citep{Asseo1998,Luo2001}.

Note that the beam and background temperatures of the streaming plasma
depend on the bunch interaction model.
In the~\citet{Usov1987} model, the background and beam temperatures are
not constrained.
As long as the bunches overlap~\citep{Usov2002}, no specific
temperature is necessary for the interaction between the bunches, such that different
temperatures are conceivable.
\citet{Weatherall1994}, e.g., assumed a highly relativistic background
temperature ($\rho_0 = 0.019$) and a cold beam ($\rho_1 = 60$).
\citet{Rafat2019b,Rafat2019a} derived the plasma dispersion properties for an arbitrary ratio between
the beam and background temperatures.
They assumed, however, equal  temperatures of both components.

\citet{Weatherall1994,Rafat2019b} assumed a weak beam, i.e.,
$r_\mathrm{n} = n_1 / (n_0 \gamma_\mathrm{b}) \ll 1$.
The use of this limit facilitates the analytical calculation of growth
rates since the wave mode
branches are determined by the background plasma while the
growth rates can be estimated by using the unstable beam
distribution.
In the \citet{Usov2002} model, the beam density is in principle
variable and can have a value equal to or larger than the background density.

\citet{Manthei2021} solved the linear dispersion relations in order
to estimate the instability growth rates and predict the unstably 
generated wave modes for a broad range of pulsar pair plasma 
parameters, re-analyzing both the weak-beam 
and the strong-beam models and validated their analytical predictions   
by means of Particle-in-Cell (PIC) code kinetic simulations.
They found that for $\rho_0,\rho_1 \geq 1$, $r_\mathrm{n} \geq 10^{-3}$
and $13 < \gamma_\mathrm{b} < 200$, the growth rates $\omega_\mathrm{i}$
of streaming instabilities are sufficiently large
to generate an instability, revealing growth
time scales $1 / \omega_\mathrm{i}$ much shorter than the time of
mutual bunch interaction $\Delta t = 10\,\mu$s.
The growth rates depend on the local plasma frequency,
providing a growth rate threshold $\omega_\mathrm{i,thr} / \omega_\mathrm{pe} = 2.8 \times 10^{-4}$
for $\omega_\mathrm{pe} = 3.6 \times 10^9$~s$^{-1}$
and $\omega_\mathrm{i,thr} / \omega_\mathrm{pe} = 1.3 \times 10^{-6}$
for $\omega_\mathrm{pe} = 7.9 \times 10^{11}$~s$^{-1}$.
This is the minimum rate required for sufficient instability
growth during the bunch interactions.

Kinetic simulations of pair plasma instabilities have also been
studied by \citet{Silva2003,Tautz2007,Cottrill2008,Lopez2014,Dangelo2015,Lopez2015,Shukla2015},
mostly initialized with three-dimensional particle distribution functions.
\citet{Shalaby2017,Shalaby2018} used one-dimensional distributions focusing on the
instability growth and on the correct resolution of the wave vector space.
\citet{Shalaby2017}, however, studied relativistic distributions only for $\gamma_\mathrm{b} \leq 4$.
Though \citet{Shalaby2018} used $\gamma_\mathrm{b} = 100$, their investigations
focused on the effects of an inhomogeneous background density.
They found the saturation level of the unstably growing electrostatic waves
but no Langmuir turbulence or solitary waves.
\cite{Jao2018} studied the formation of electrostatic Alfv\'{e}n waves in a cold
streaming relativistic pair plasma.

Despite of decades of research, it is still necessary to solve a
number of questions about the possibility of pulsar radio emissions
due to the formation of solitary plasma waves~\citep{Rahaman2020} 
--- the mechanism of generation of large amplitude electrostatic 
waves, the derivation of growth rates for realistic plasma parameters, 
and the question of whether the instability growth is sufficiently fast 
to allow reaching a strongly nonlinear stage.

\citet{Melrose2020a} have discussed two types (stages) of
the formation of solitary waves.
To our knowledge, their formation has, however, not been simulated 
by kinetic PIC-codes yet.
Their properties and consequences are, therefore, not known, yet.
We now have investigated the process leading to the second 
stage of solitary wave formation by means of kinetic numerical 
simulations.

In order to understand the consequences of the formation of solitary
waves for the pulsar radio emission, we have studied the long-time 
evolution of the underlying streaming instabilities in typical 
relativistic pair plasmas of pulsar magnetospheres.
Our simulations cover the growth of electrostatic waves, their 
saturation as subluminal L-mode waves,
the formation of superluminal L-mode waves and solitons.
We identify the parameter range for which solitary waves are formed.
By calculating the saturation levels of the electrostatic wave energy, we
estimate the energy densities of the possible pulsar radio emission via this mechanism.

This paper is structured as follows.
In Section~\ref{sec:methods}, we describe the investigation methods 
we used, i.e. our Particle-in-Cell code numerical simulations, their 
initial setup and boundary conditions.
In Section~\ref{sec:results}, we present detailed analyses of the
simulation results.
We estimated the resulting electromagnetic wave energies and 
discussed them in Section~\ref{sec:discuss} and summarized our
results in Section~\ref{sec:conc}. 
In two appendices we discuss additional issues, in Appendix \ref{AppendixA} 
the properties of the L-mode waves, we found and in 
Appendix \ref{AppendixB} the stability and convergence 
of our simulation results.

\section{Methods} 
\label{sec:methods}
For our study of the formation of solitons in a relativistic pair
pulsar plasma, we use fully-kinetic Particle-in-Cell (PIC)
numerical simulations carried out with the code ACRONYM \citep{Kilian2012}.
In order to efficiently suppress numerical Cherenkov radiation,
which is a numerical
effect that appears in relativistic PIC simulations due to the incorrect light
wave propagation, the code uses two special features.
The first one is the Cole-K\"arkk\"ainen (CK) \citep{Cole1997,Karkkainen2006}
non-standard finite difference scheme (NSFD) for the Maxwell field solver.
The second feature is a recently developed interpolation scheme
(also known as shape function
or weighting function) that is used
to deposit the particles' current density onto the grid and to get
the electromagnetic fields at the particles' positions. This scheme is called
``weighting with time-step dependency'' (WT) scheme  \citep{Lu2020}.
We used its 4th order version. The use of such an arbitrary interpolation
scheme is compatible with the use of the
charge conserving current deposition scheme by \citet{Esikperov2001}.
All those features do not only provide exact momentum and
charge conservation but also suppress the numerical Cherenkov radiation
to a very high degree and conserve the total energy in the system very well.

We conduct a series of simulations to sufficiently cover the
parameter space. Each 1D simulation is along the $x$-direction.
That is also the direction of the magnetic field.
However, the latter is not included in the simulation,
as there is no interaction of the electric and magnetic
fields when both vectors are along the same $x$-direction.
We use a time step $\omega_\mathrm{p} \Delta t = 0.03519$
and a normalized grid cell size $\Delta x = 0.07108\,d_\mathrm{e}$,
where $\omega_\mathrm{p}$ is the total plasma frequency of all species.
Each simulation is $10^4$ grid cells ($L \approx 711\, d_\mathrm{e}$) long.
This length gives sufficient resolution of the unstable waves
in Fourier space ($\Delta k c / \omega_\mathrm{p} = 8.8 \times 10^{-3}$).
Each simulation runs for a total time $2 \tau_\mathrm{sat}$,
where $\tau_\mathrm{sat}$ is the simulation saturation time.
If solitons do not appear until 600000 times steps ($\omega_\mathrm{p} t \approx 20000$),
the simulation is stopped. 
The maximal resolution in frequency is $\Delta \omega / \omega_\mathrm{p} = 4.7 \times 10^{-5}$.
It is inversely proportional to the total simulation time, so that $\Delta \omega$ 
is larger for our shorter simulations.
Periodic boundary conditions are applied.
With these simulation parameters, the generated solitons are stable,
and their properties are not significantly influenced by additionally
increasing the number of particles or the numerical resolution
(Appendix~\ref{AppendixB}).

In total, we use four particle species --- two populations
of electrons and two of positrons.
They are divided between the background composed of electrons and positrons,
and the beam that is also composed of the same species.
The particles follow the Maxwell-J\"uttner velocity distributions
described by Equations~\eqref{eq1}--\eqref{eq3}.
They are generated using the rejection method that initially
selects particles from a much wider velocity interval that is
expected from their distributions.
We tested our resulting distribution against the analytical function,
by increasing he number of particles in the simulation box by a factor of $100$.
We found  a good agreement since the resulting distribution approaches the analytical solution,
even for the tail of the distribution.
For our simulations, we use two number density ratios between the background and beam particles:
Either keeping $r_\mathrm{n} = \mathrm{const}$
or choosing $n_0 = n_1$. In the case $r_\mathrm{n} = \mathrm{const}$,
we used $10^4$ background particles per grid cell.
The number of beam particles changes with $\gamma_\mathrm{b}$
to ensure that $r_\mathrm{n} = \mathrm{const}$.
In the case $n_0 = n_1$, $10^3$ particles for the background
and $10^3$ particles for the beam are used independently on $\gamma_\mathrm{b}$.
The temperature of the background is selected to be equal
to the beam temperature in all our simulations, i.e, $\rho_0 = \rho_1$.

We carry out a set of simulations for a variety of  beam velocities,
background and beam temperatures, and beam-to-background density ratios.
We vary the beam velocity in the range
of Lorentz factors between $\gamma_\mathrm{b} = 20-200$.  These parameter ranges
cover the transition in the parameter space between
the cases without and with L-mode solitons.

\citet{Arendt2002} determined the temperature of the generated bunch and other properties  by means of numerical simulations of the pair creation process in pulsar magnetospheres.
They found that the secondary beam features a typical temperature of
$\rho \sim 1$ for $B_\star = 10^{12}$~G, where $B_\star$ is the magnetic field
strength at the height where the photon ($\gamma$-ray) seed emission occurs.
For $B_\star = 10^{13}$~G the temperature varies by almost two orders of magnitude.
The streaming instability is assumed to be formed by the slowest
and fastest particles that are released from this hot bunch.
However, the temperature (given by the velocity spread),
of the released particles is lower
than that in the original bunch \citep{Rahaman2020}.
The temperature varies in the interval $\rho_0 = \rho_1 = 1 - 3.33$.
It represent the transition for the formation of superluminal L-mode solitons.
Although for temperatures $\rho < 1$ solitons can also form, 
\citet{Manthei2021} found that the instability growth rate is too low if the beam velocity is large enough.

Note that all quantities are in CGS units.

\section{Results} \label{sec:results}
As we are dealing with the dispersion properties
of waves modes in relativistic plasmas, the naming convention
of waves and properties is slightly different from that used
in non-relativistic plasmas, where ``classical'' Langmuir
 waves are present \citep{Rafat2019c}.
In the context of this paper, we distinguish between 
\textit{subluminal waves}, \textit{subluminal solitons}, \textit{superluminal waves},
and \textit{superluminal solitons}.
All the present waves are longitudinal waves along the magnetic field lines
in relativistically hot pair plasmas.
We denote all electrostatic waves located at the superluminal part of the L-mode (the relativistic form of the Langmuir branch) dispersion branch as \textit{superluminal waves}.
\textit{Superluminal solitons} are
stable large-amplitude waves located at the superluminal part of the L-mode
dispersion branch.
These types of solitons are stable for a long time in the course
of a simulation run,
and their typical size is larger than the size of the associated
electron-positron phase space holes.
Both superluminal L-mode waves and solitons are generated only for
a specific range of plasma parameters.
\textit{Subluminal solitons} are connected with electrostatic
waves located at the unstable subluminal L-mode dispersion branch.
They are also associated with electron-positron phase space holes,
and are stable for a long time. 
These subluminal solitons cause electrons and positrons density waves that are
shifted in phase in such
a way that the total particle density is approximately constant.
This type of solitons appears in all our simulation runs.

\subsection{Saturation Types}
\begin{figure*}[!ht]
    \gridline{\fig{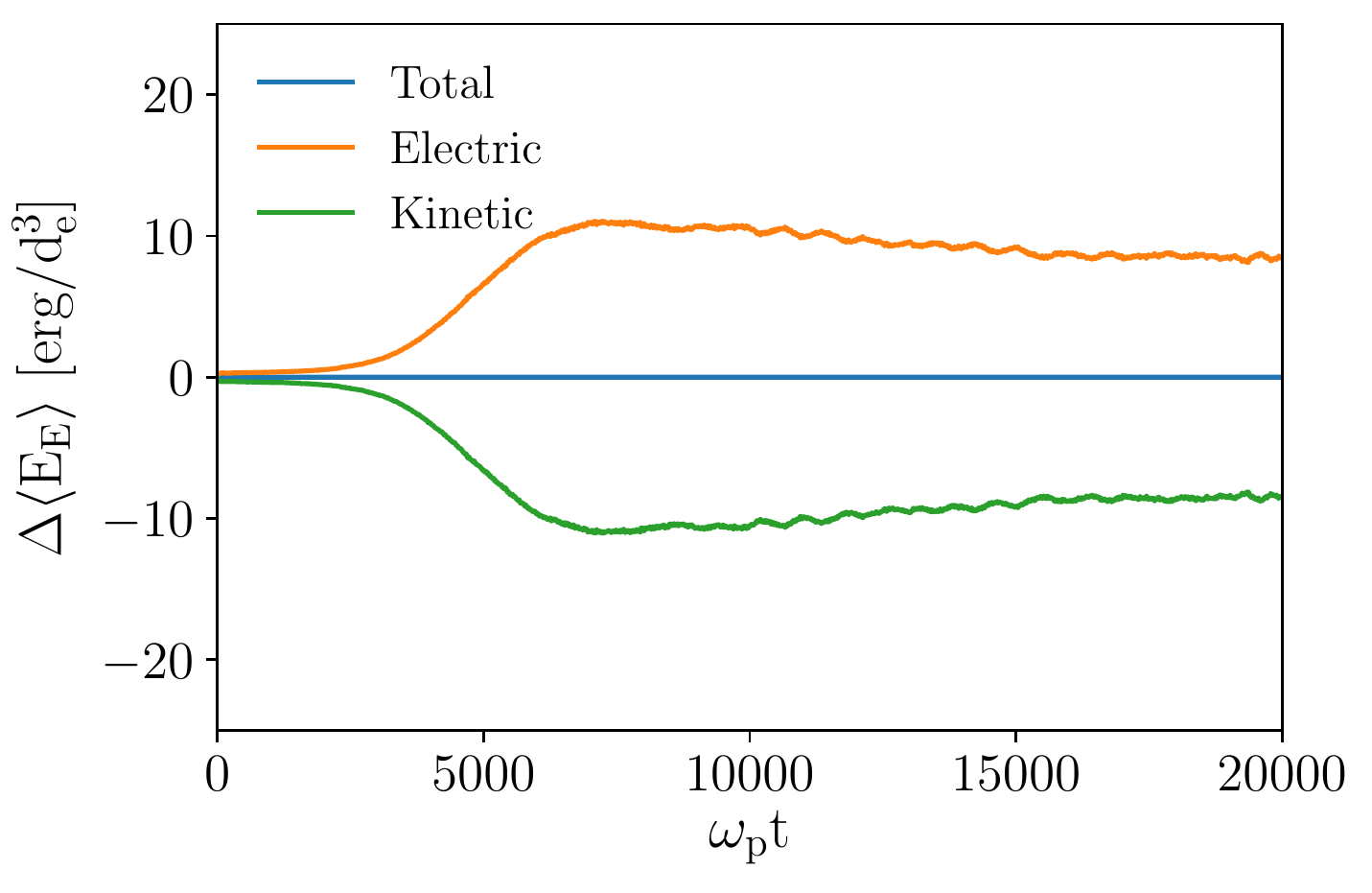}{0.333\textwidth}{(a) $\rho_0 = \rho_1 = 1, \gamma_\mathrm{b} = 26$. Saturation Type 1.}
              \fig{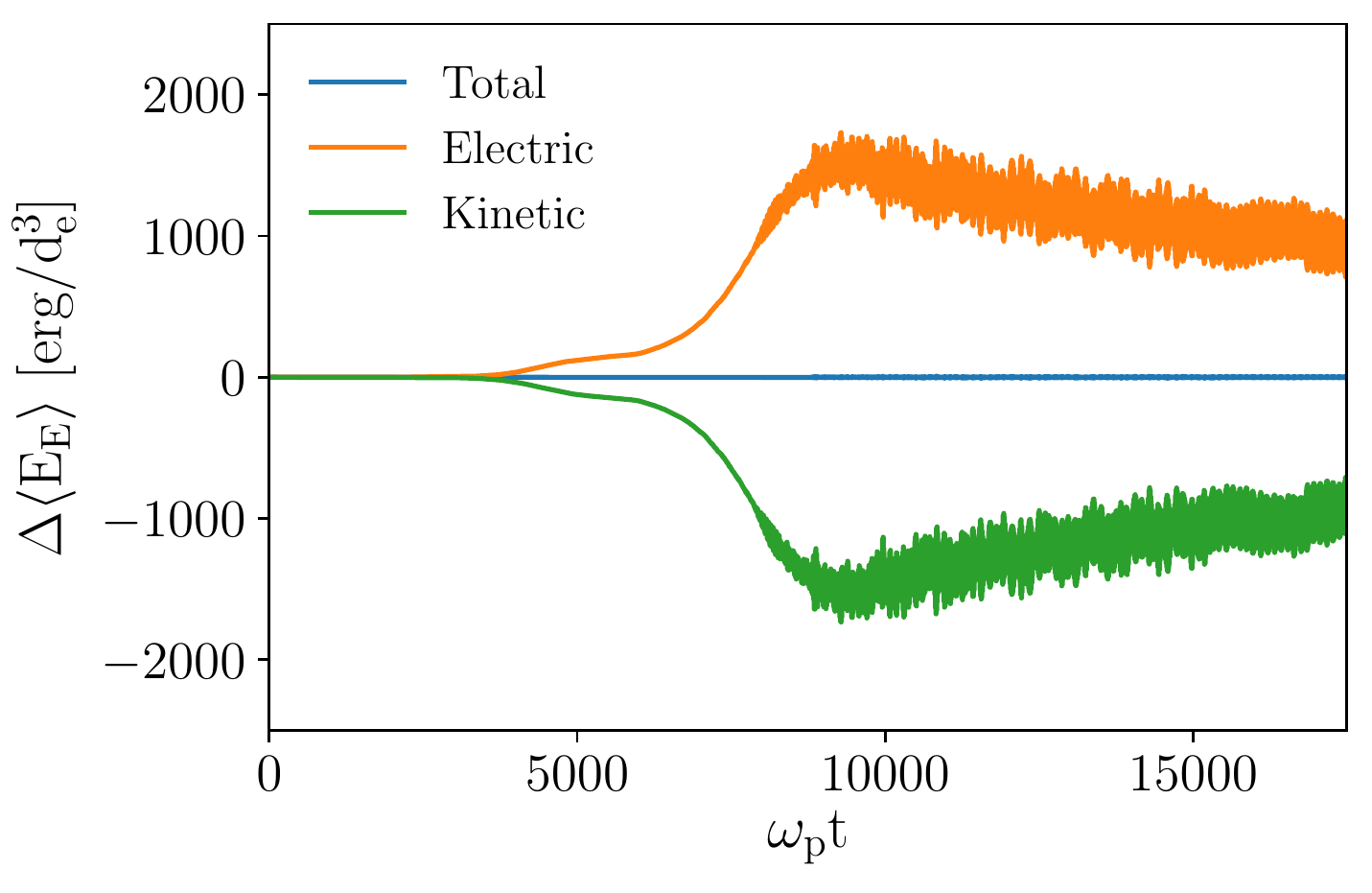}{0.333\textwidth}{(b) $\rho_0 = \rho_1 = 1, \gamma_\mathrm{b} = 103$. Saturation Type~1 followed by Type~2.}
              \fig{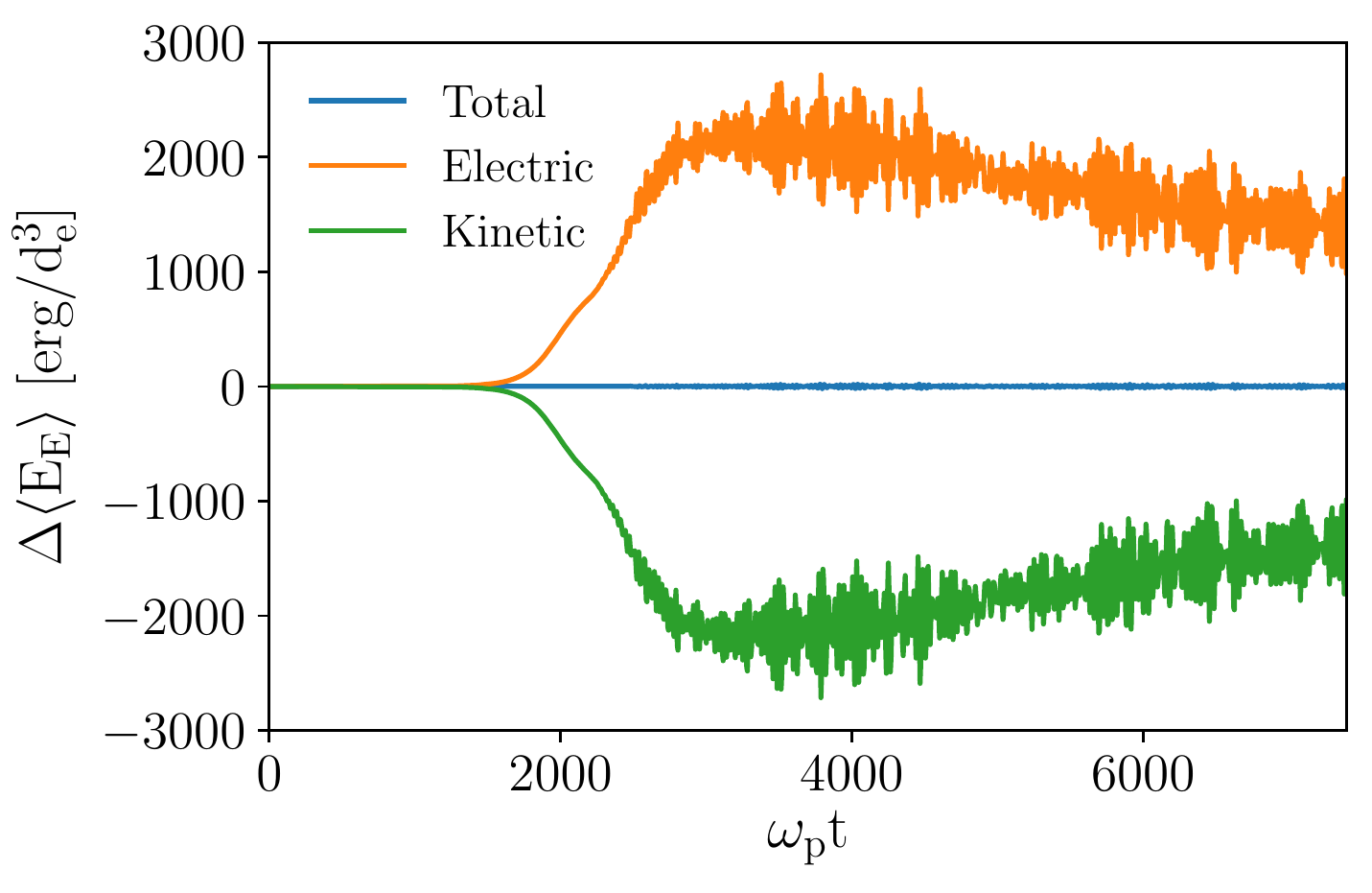}{0.333\textwidth}{(c) $\rho_0 = \rho_1 = 3.33, \gamma_\mathrm{b} = 60$. Immediate saturation Type 2.}
          }
    
    \caption{Evolution of the total, kinetic,
    and electric energy density difference from the initial state
    for three selected cases with different saturation types. 
    Simulations differ in initial temperatures $\rho_0$, $\rho_1$,
    and the beam Lorentz factor $\gamma_\mathrm{b}$;
    $r_\mathrm{n} = 10^{-3}$ is kept constant.}
    \label{fig1}
\end{figure*}
We conducted several simulations in the parameter space mentioned above.
Our simulations can be sorted into three \emph{categories}
depending on the behavior of the electric field after saturation.
We number them as follows:

\begin{enumerate}
 \item Simulations with an initial exponential energy increase,
 followed by saturation. The electrostatic energy
 \textit{remains constant} for the whole remaining computing time. 
 \item The same initial energy exponential evolution
 with saturation as category 1.
 \textit{After saturation}, however,
 the energy begins to \textit{grow again} and finally saturates
 for a second time at a much higher energy level than in the first case.
 Moreover, the electrostatic energy after the second
 saturation exhibits relatively large oscillations.
 \item The electrostatic energy evolution shows
 a quicker exponential increase and constant saturation
 very similar to the first type.
 However, this first saturation has the same \textit{properties as the second saturation type}
 in the second category of simulations, i.e.,
 high energy saturation level as well as energy oscillations later on.
\end{enumerate}
Overall, these three categories manifest two different
saturation types represented by the categories one and three.
Category two is the transition between them.
We denote the recognized saturation types as saturation Type~1
(lower electrostatic energy density with smooth evolution)
and saturation Type~2 (higher energy density with oscillating evolution).
As an illustration, we select three representative cases 
for every category of simulations:
\begin{enumerate}
\item $\rho_0 = \rho_1 = 1, \gamma_\mathrm{b} = 26$.
\item $\rho_0 = \rho_1 = 1, \gamma_\mathrm{b} = 103$.
\item $\rho_0 = \rho_1 = 3.33, \gamma_\mathrm{b} = 60$.
\end{enumerate}
Figure~\ref{fig1} presents the time evolution of the mean total,
electrostatic, and kinetic energy density for these three cases.
The initial energy density is subtracted.
The energy densities are normalized as energy per volume
of a cube of size $d_\mathrm{e}^3$, with $d_\mathrm{e}=c / \omega_\mathrm{p}$  the skin depth. 
In Case~1, the electrostatic energy exponentially increases
with a low growth rate until it saturates at $\omega_\mathrm{p}t \sim 7500$.
In Case~2, the saturation occurs earlier at $\omega_\mathrm{p}t \sim 5500$.
From $\omega_\mathrm{p}t \sim 7000$ on,
the energy increases again and saturates again at a much
higher energy density at $\omega_\mathrm{p}t \sim 9000$.
Case~3 shows only one energy increase with
a growth rate higher than in the two previous cases;
the saturation occurs at $\omega_\mathrm{p}t \sim 3000$.

\begin{figure*}[!ht]
    \gridline{\fig{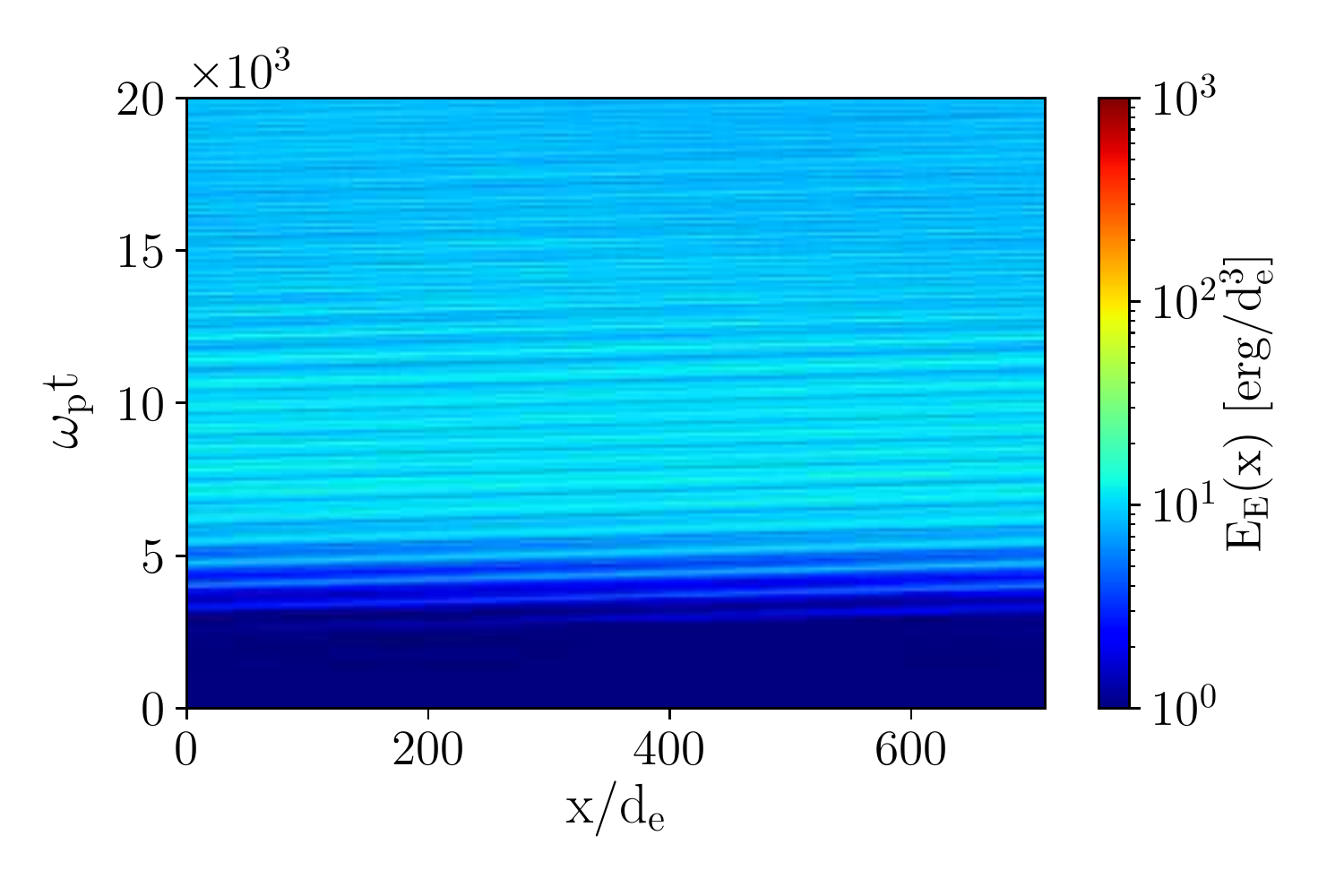}{0.333\textwidth}{(a) $\rho_0 = \rho_1 = 1, \gamma_\mathrm{b} = 26$.}
              \fig{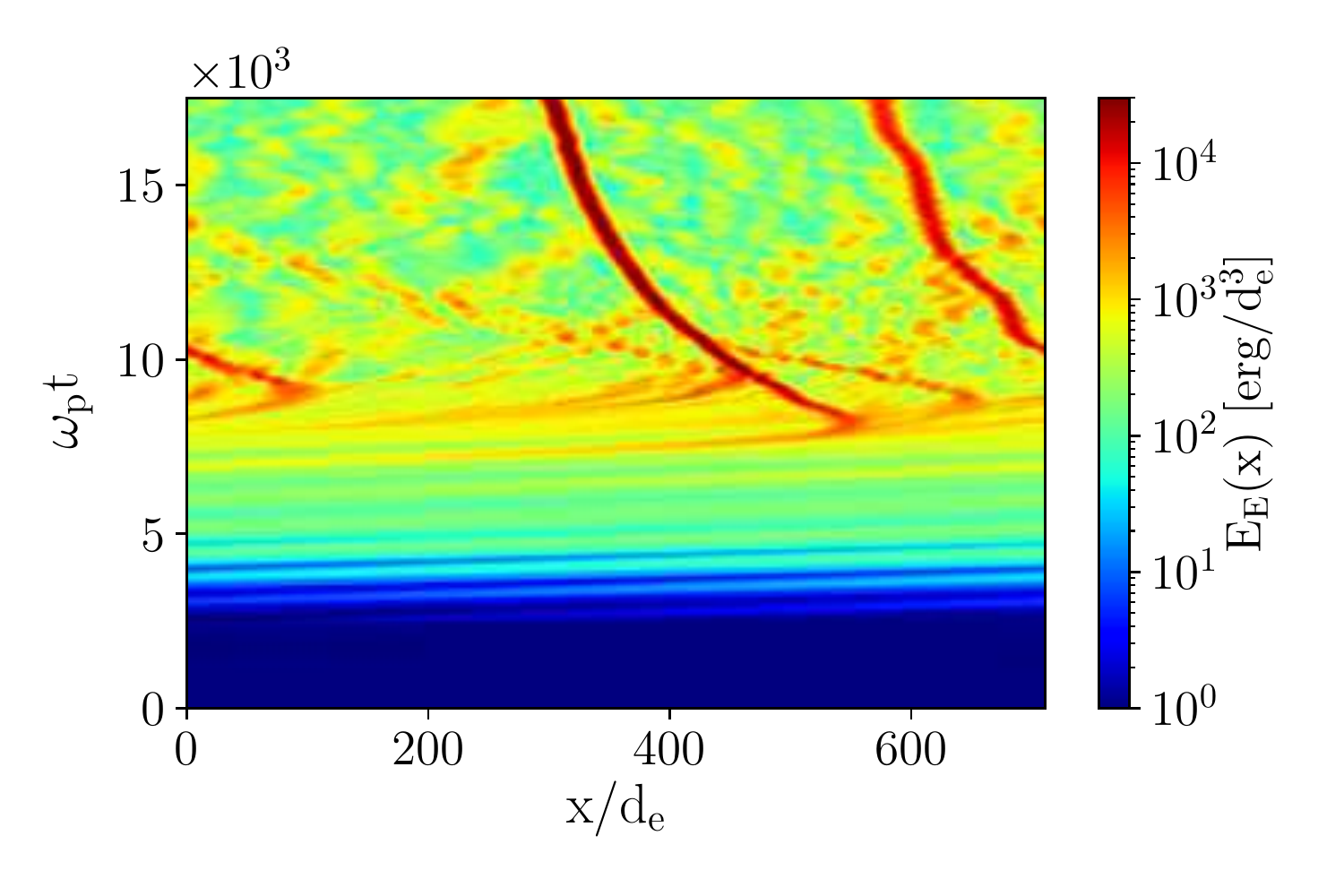}{0.333\textwidth}{(b) $\rho_0 = \rho_1 = 1, \gamma_\mathrm{b} = 103$.}
              \fig{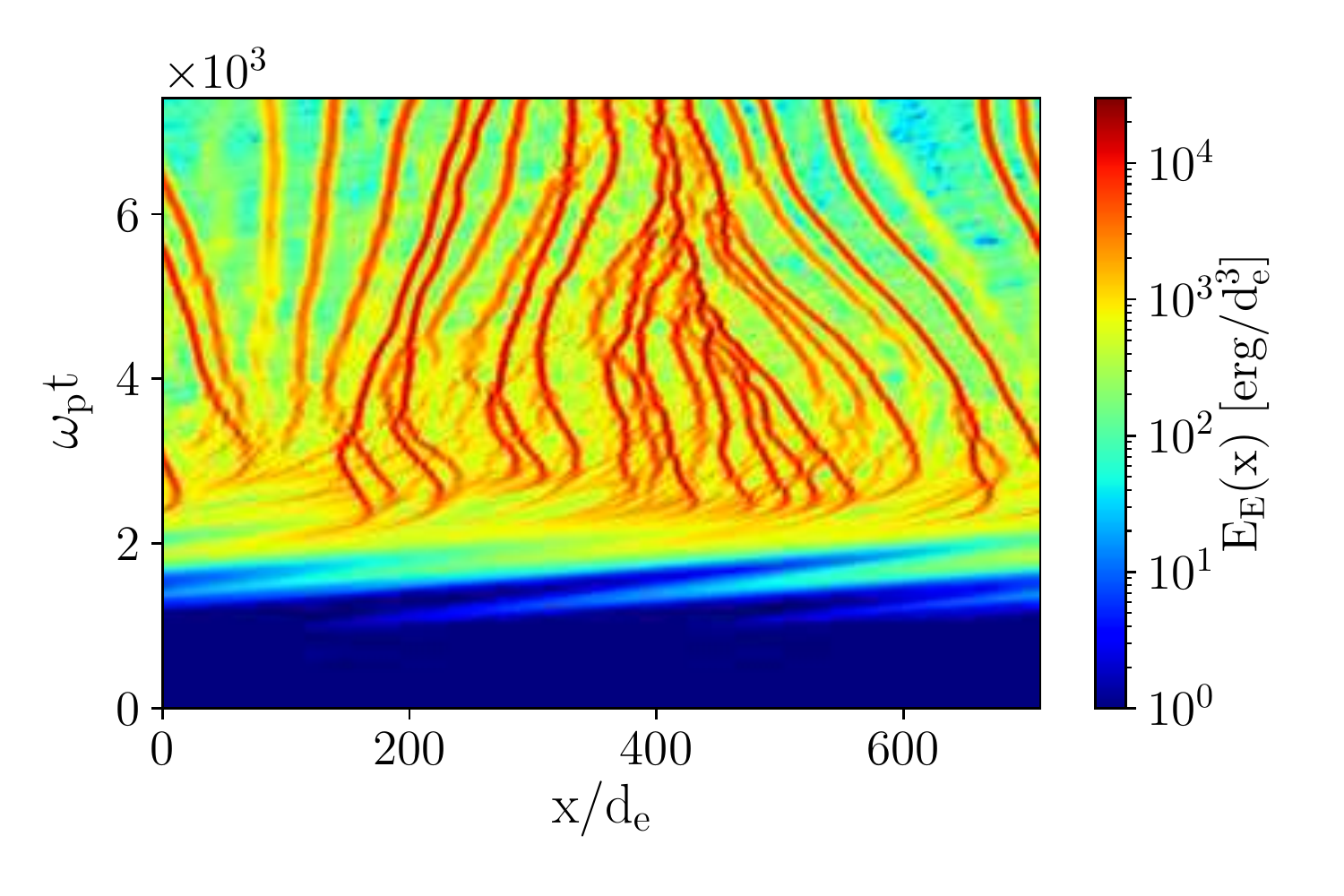}{0.333\textwidth}{(c) $\rho_0 = \rho_1 = 3.33, \gamma_\mathrm{b} = 60$.}
             }
    \caption{Evolution of the electrostatic energy density. The color scale is different between figures. Same simulations as in Figure~\ref{fig1}.}
    \label{fig2}
\end{figure*}

\begin{figure*}[!ht]
    \gridline{\fig{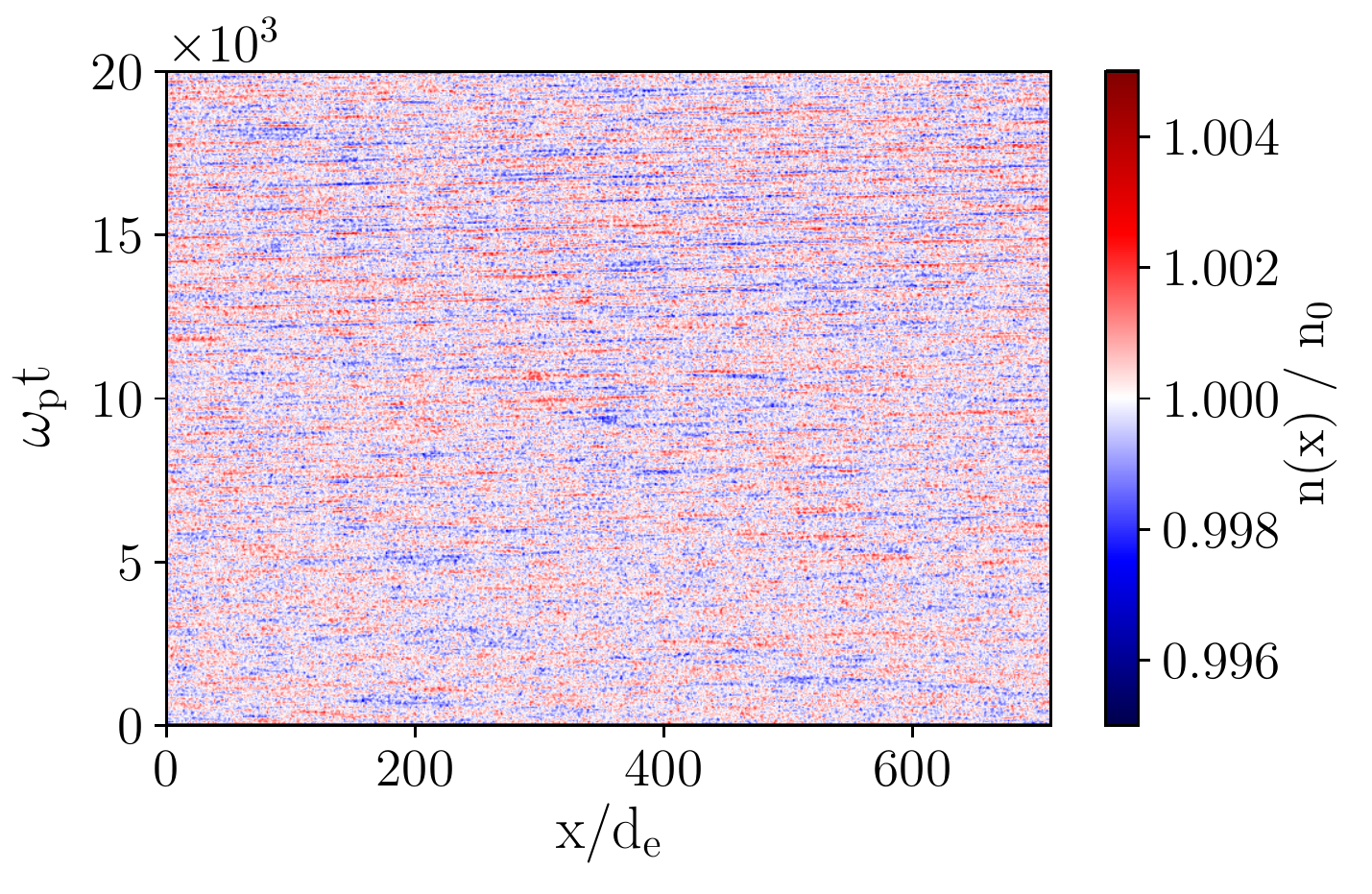}{0.333\textwidth}{(a) $\rho_0 = \rho_1 = 1, \gamma_\mathrm{b} = 26$.}
              \fig{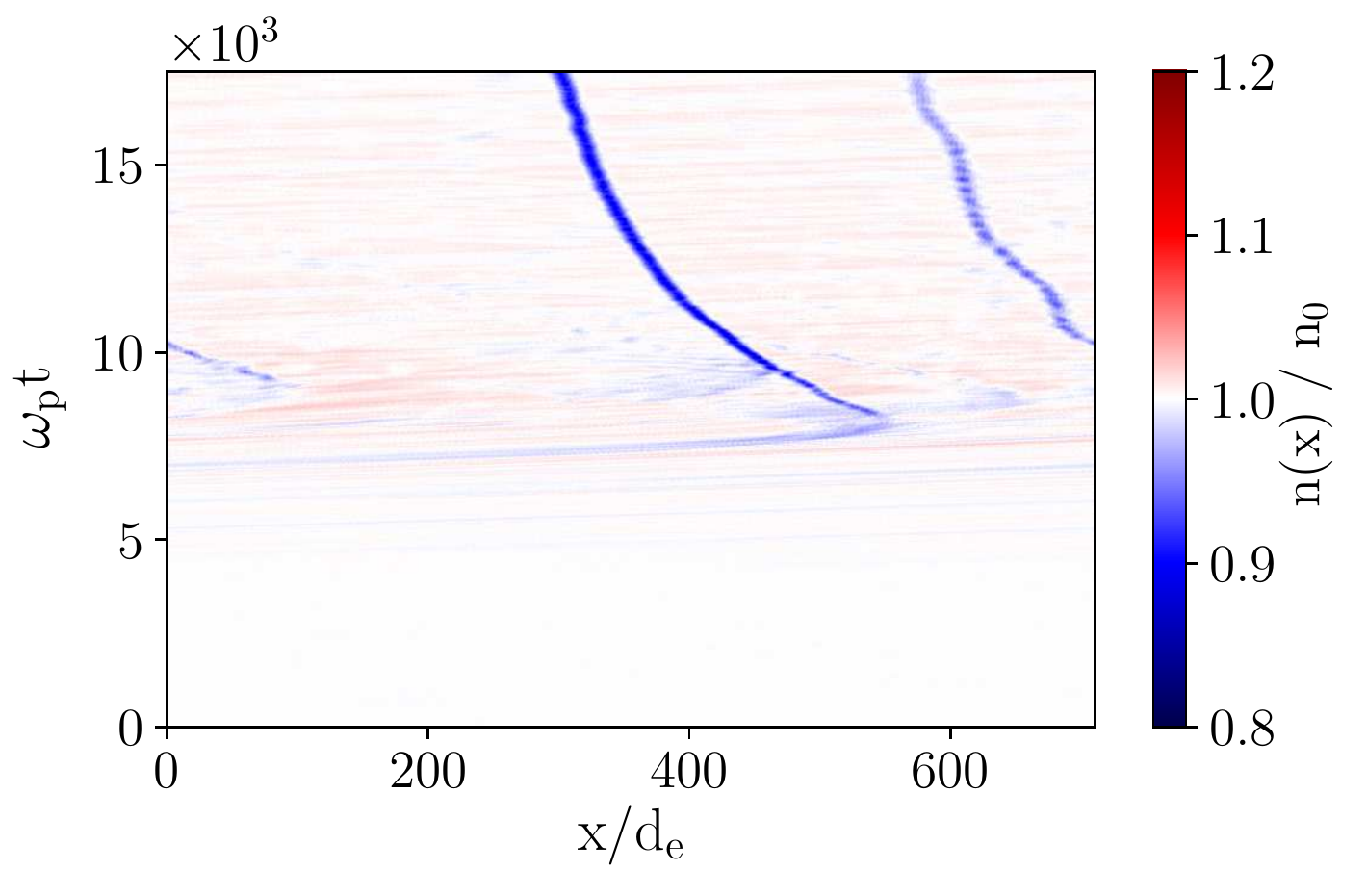}{0.333\textwidth}{(b) $\rho_0 = \rho_1 = 1, \gamma_\mathrm{b} = 103$.}
              \fig{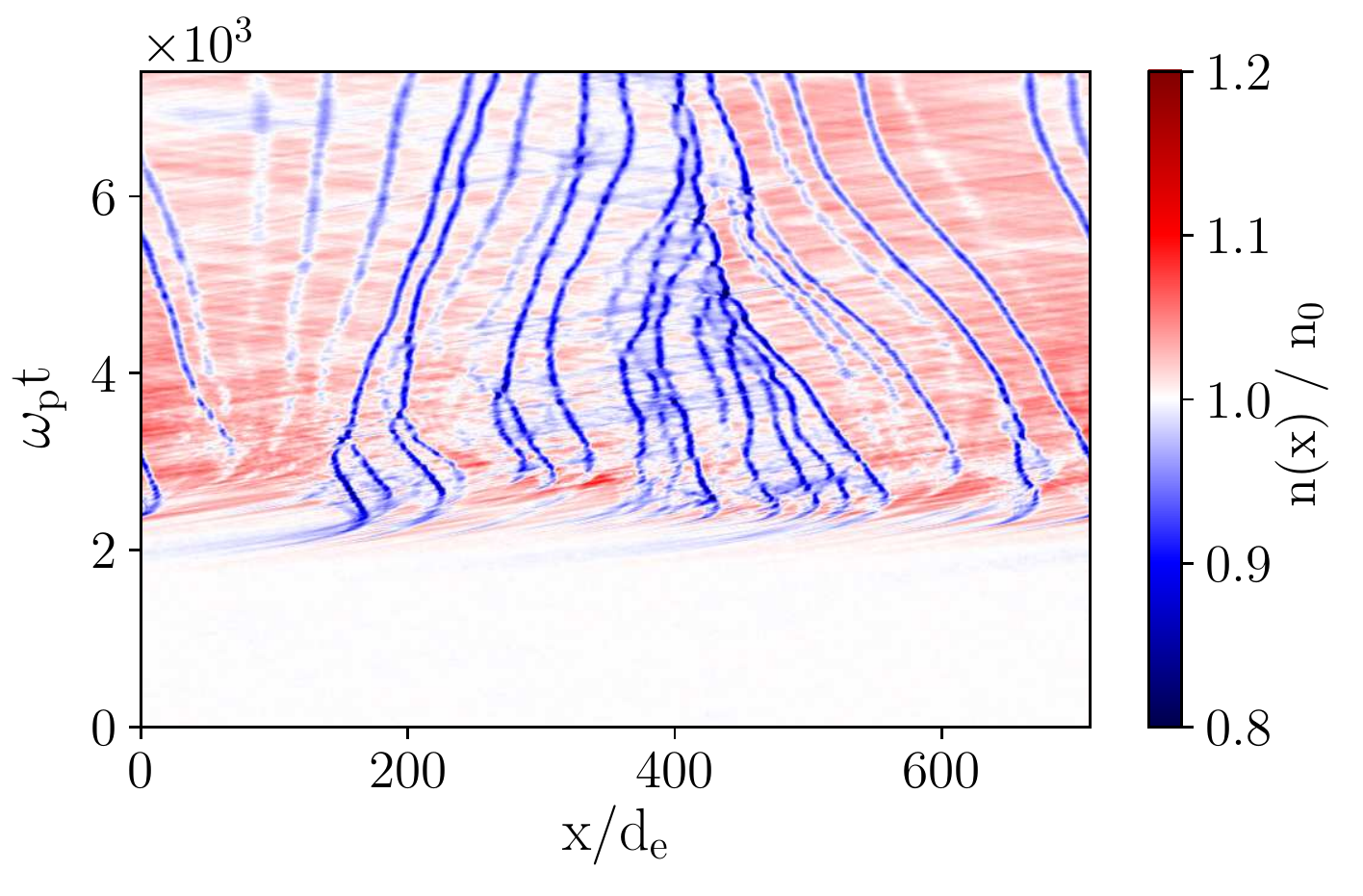}{0.333\textwidth}{(c) $\rho_0 = \rho_1 = 3.33, \gamma_\mathrm{b} = 60$.}
             }
    \caption{Evolution of the total particle density normalized to the initial density. The color scale is different between figures. Same simulations as in Figure~\ref{fig1}.}
    \label{fig13}
\end{figure*}

Figure~\ref{fig2} shows the time evolution
of the energy density profile along the $x$-direction.
The color-coded contour plots are displayed with different ranges
of a logarithmic-scale colorbar in order to cover the wide range of generated energies.
In Case~1 (Figure~\ref{fig2}a), waves are
spatially uniformly distributed along the whole simulation domain.
For the larger beam velocity of Case 2 (Figure~\ref{fig2}b),
waves are uniformly distributed during
both the initial growth stage and the first saturation (Type~1).
During the second growth, waves begin to aggregate at some locations.
Initially, they have a group velocity close to the speed of light.
Eventually, they slow down until $\omega_\mathrm{p}t \sim 8000$
when the second saturation (Type~2) occurs.
Several aggregated large-amplitude electrostatic waves,
which we identified as solitons, appear at specific positions.
In the simulation reference frame, those solitons have both positive and negative group velocities.
Solitons mutually interact and often merge.
For the Case~3 with a lower background and beam temperature (Figure~\ref{fig2}c),
solitons already appear during the first saturation Type~2.
In this Case~3, the distance between solitons,
their size and oscillation amplitude are all smaller than those for the Case~2.

Figure~\ref{fig13} shows the time evolution
of the total particle density profile along the $x$-direction.
The plots are displayed with different colorbar ranges.
In Case~1 (Figure~\ref{fig13}a), only density fluctuations of low amplitude
waves are present.
For the larger beam velocity of Case 2 (Figure~\ref{fig13}b),
density waves begin to aggregate at the locations of enhanced electrostatic waves during the second growth period.
For the Case~3 (Figure~\ref{fig13}c),
density depressions also appear at the same positions as the electrostatic waves.
Moreover, as particles are pushed out of the solitons,
the regions between them have mostly higher density than the initial one.

\begin{figure*}[!ht]
    \gridline{\fig{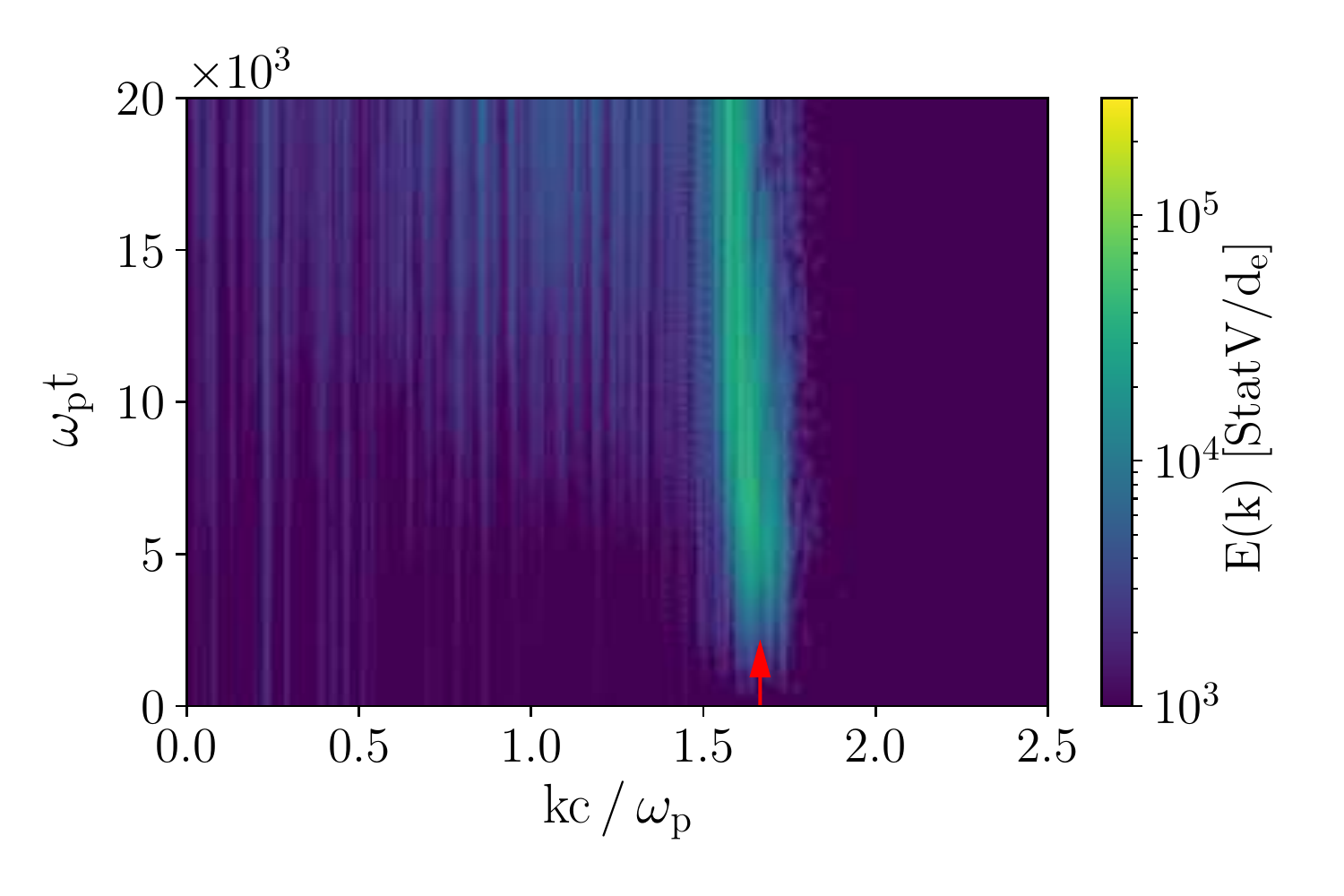}{0.333\textwidth}{(a) $\rho_0 = \rho_1 = 1, \gamma_\mathrm{b} = 26$.}
              \fig{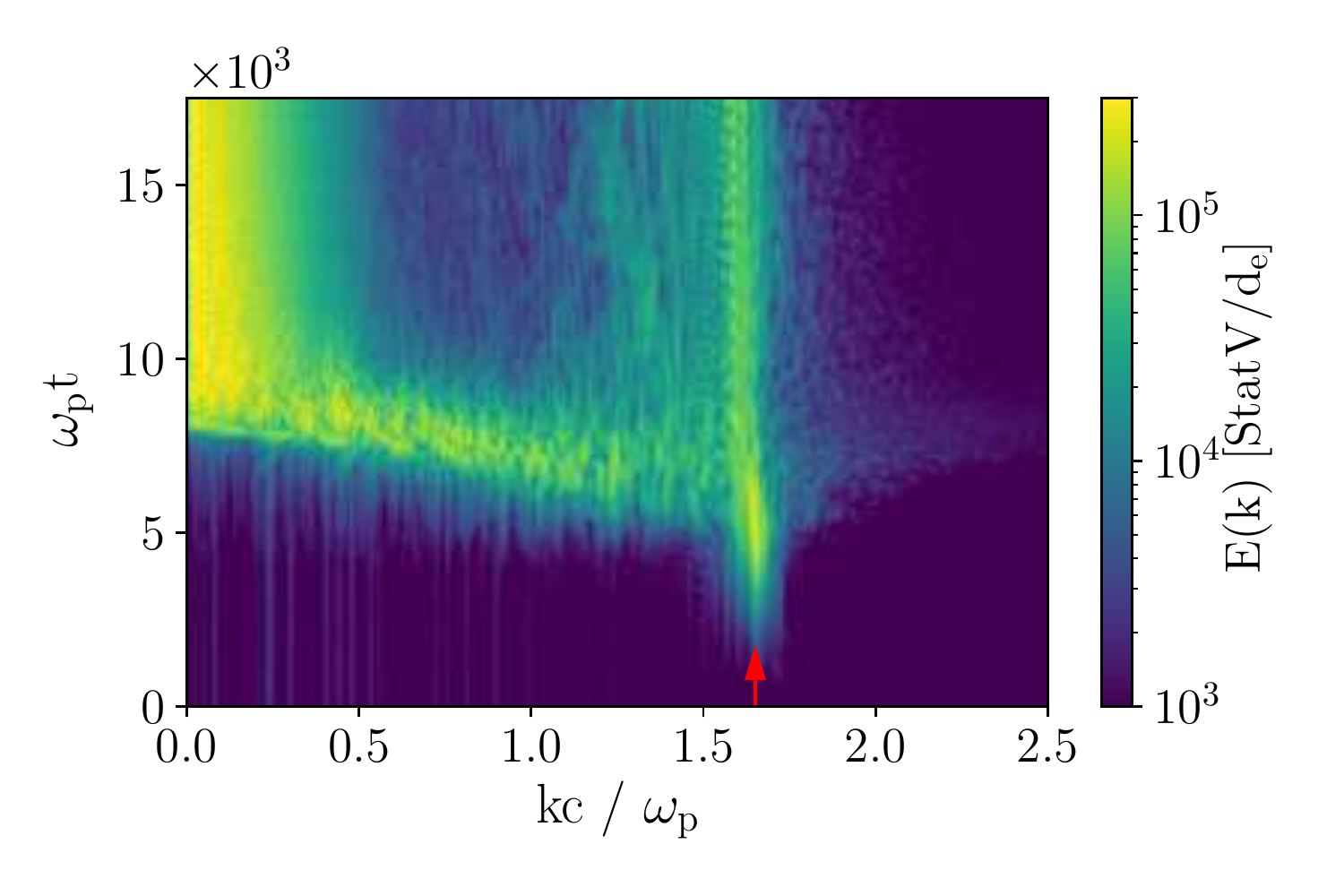}{0.333\textwidth}{(b) $\rho_0 = \rho_1 = 1, \gamma_\mathrm{b} = 103$.}
              \fig{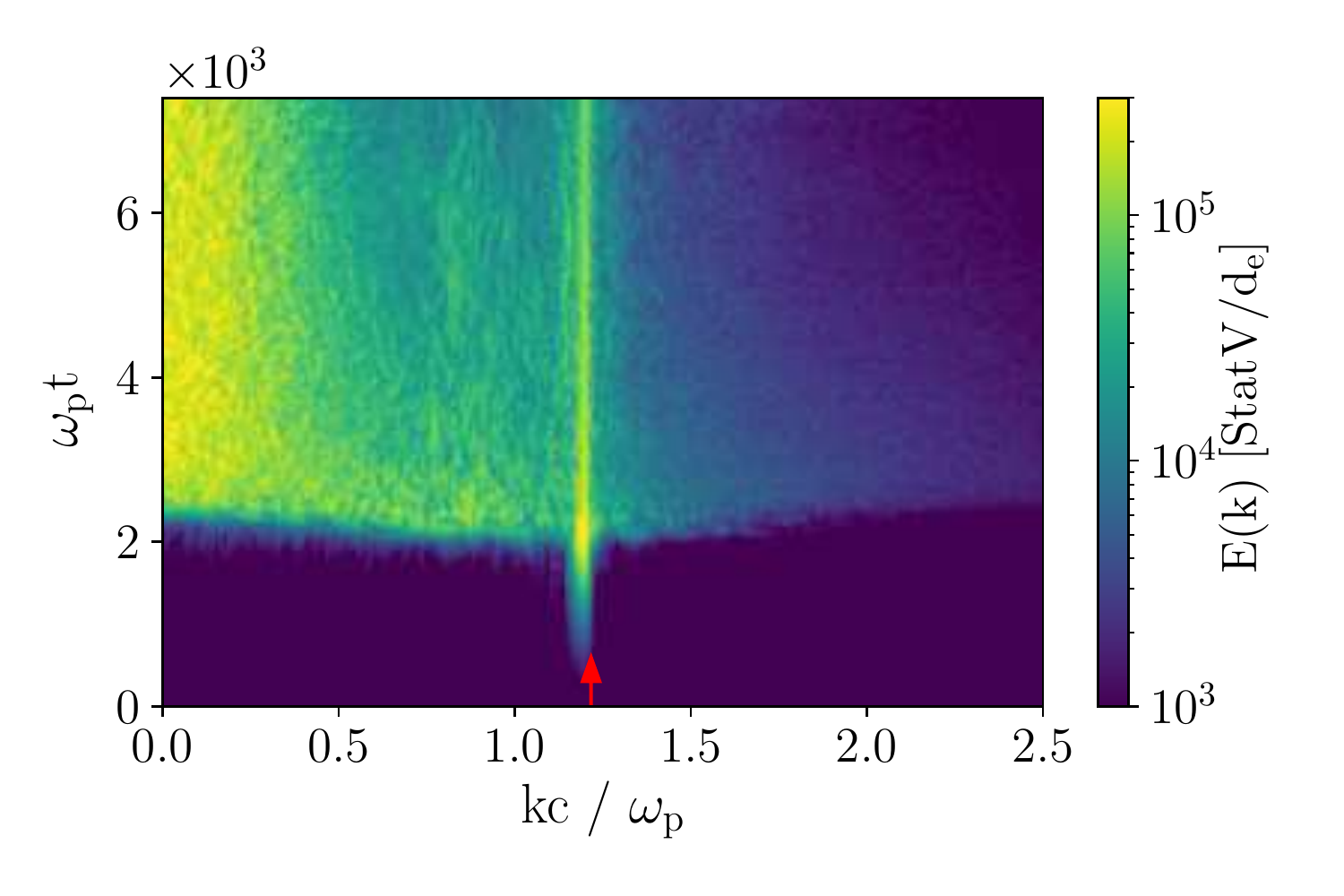}{0.333\textwidth}{(c) $\rho_0 = \rho_1 = 3.33, \gamma_\mathrm{b} = 60$.}
             }
    \caption{Evolution of the wave number of electrostatic waves.
    The red arrows denote the wave number of the unstable wave with
    the highest value of the growth rate that is predicted by analytical theory,
    (a) and (b) taken from \citep{Manthei2021}. Same simulations as in Figure~\ref{fig1}. }
    \label{fig3}
\end{figure*}
Figure~\ref{fig3} presents the evolution of the wave number of those electrostatic waves.
It is calculated as the Fourier transform of the intensity
of the electric field along the $x$-axis in the whole simulation domain.
In Case~1 (Figure~\ref{fig3}a), the amplitude of the unstable
electrostatic waves increases at $k c / \omega_\mathrm{p} \sim 1.65$.
As explained below, we identified those electrostatic waves as subluminal waves.
After these unstable waves saturate, their amplitude remains roughly constant.
In Case 2 (Figure~\ref{fig3}b), after the saturation of the initially
unstable wave at $k c / \omega_\mathrm{p} \sim 1.65$,
the range of unstable wave numbers broadens, and they shift toward longer wavelengths.
The amplitude of the initially saturated waves decreases and remains constant.
In Case 3 with a lower plasma temperature (Figure~\ref{fig3}c),
the waves first grow in a relatively narrow wave number range centered
around $k c / \omega_\mathrm{p} \sim 1.20$.
After they saturate, waves with a broader range of wave numbers are created,
reaching $k \sim 0$ within a much shorter time than for the previous case.
For even lower initial plasma temperatures,
the wave amplitudes increase almost immediately at all wavelengths at the saturation time.

The linear effects occur
at the beginning of the instability development.
The nonlinear effects start to influence the instability later,
when the unstable wave modes gain more energy.
In order to compare the initial (linear) evolution of the instability 
simulations with linear theory,
we analytically computed the linear growth rates as a function
of the frequency and wave number using the methods developed in \citet{Manthei2021}.
They are also summarized in the Appendix~\ref{AppendixA}.
The red arrows denote the wave number of the unstable waves
that were found to have the highest growth rates in the analytical linear theory.
It is theoretically predicted that waves with these wave numbers begin to grow first.

\begin{figure*}
    \gridline{\fig{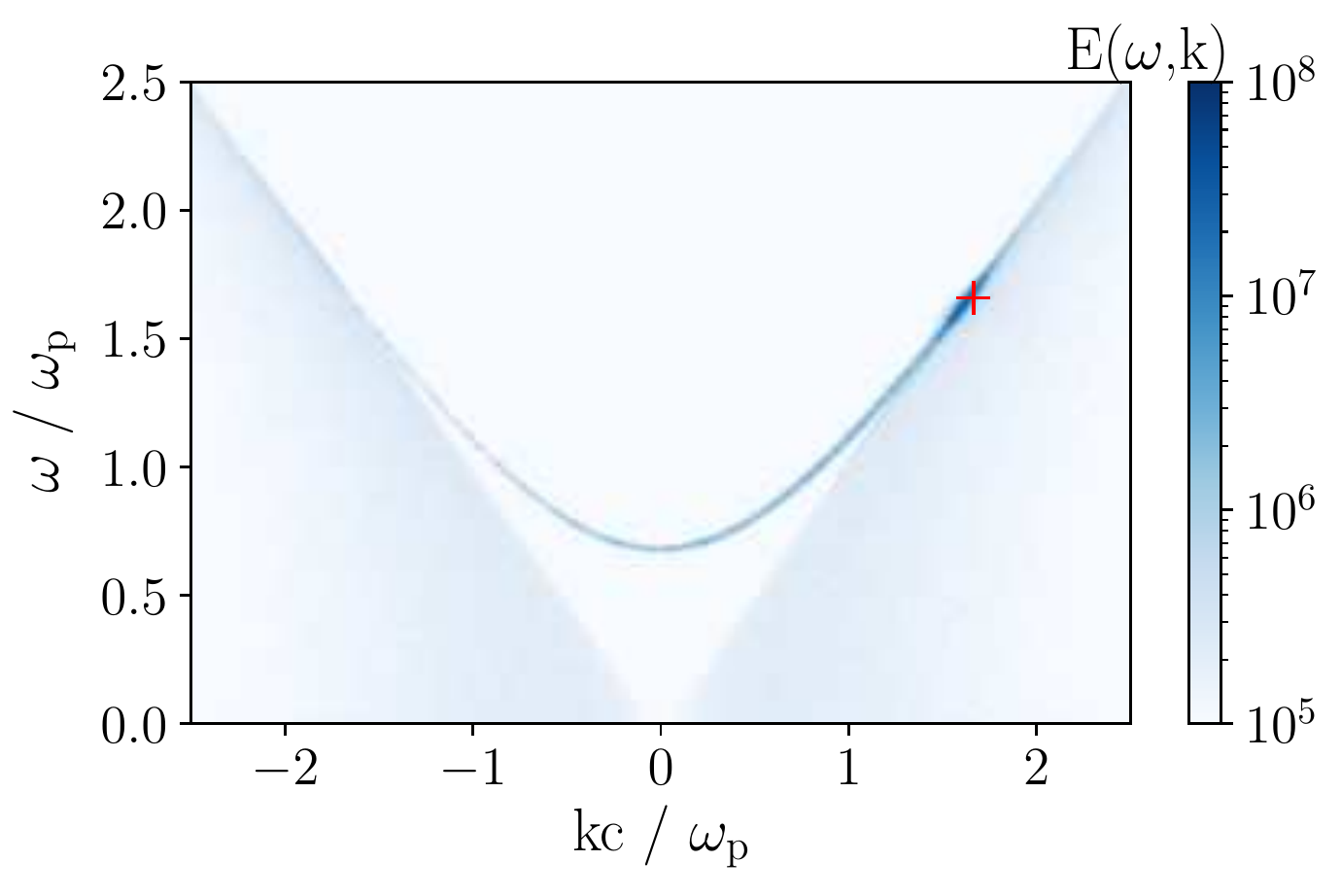}{0.333\textwidth}{(a) $\rho_0 = \rho_1 = 1, \gamma_\mathrm{b} = 26$.}
              \fig{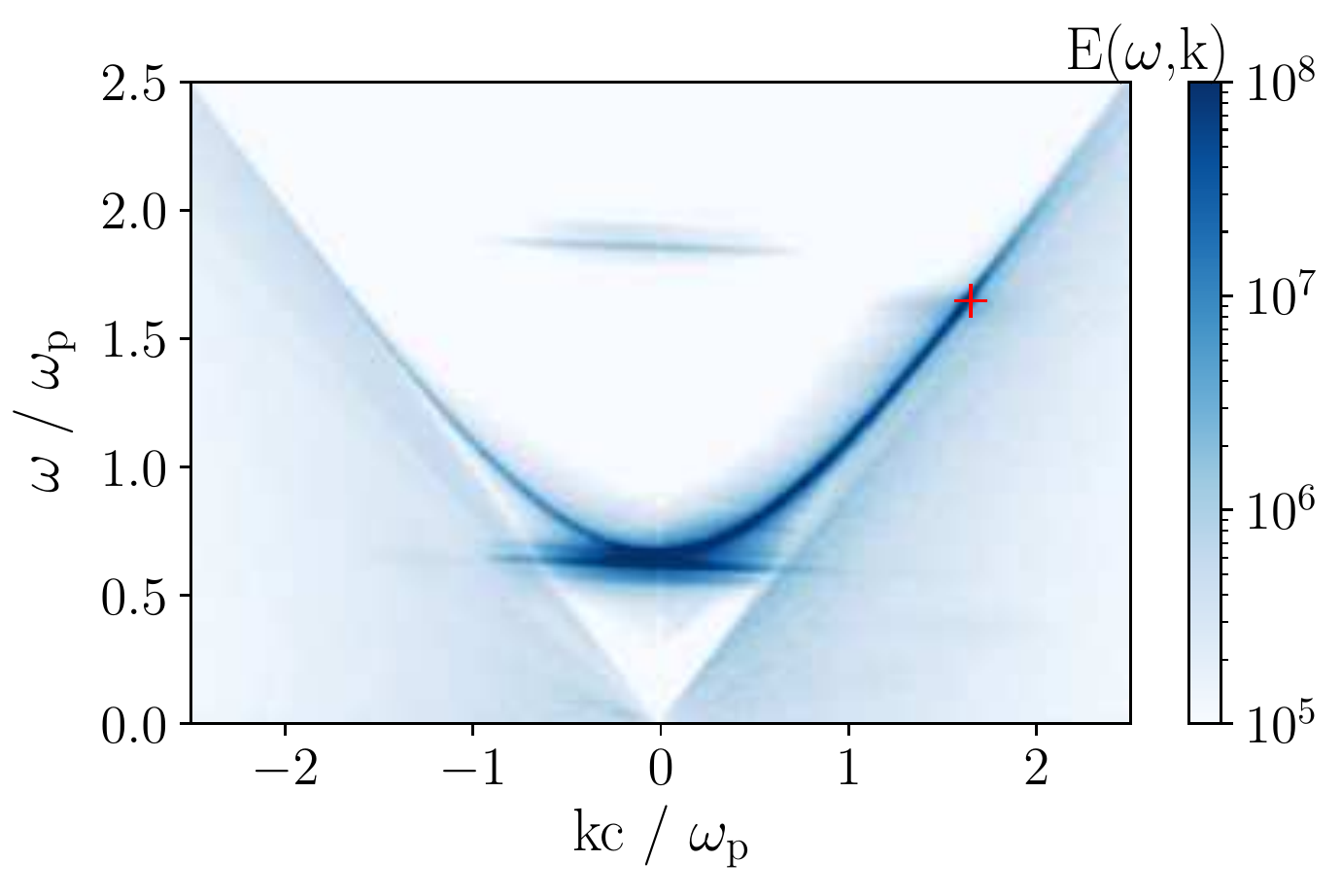}{0.333\textwidth}{(b) $\rho_0 = \rho_1 = 1, \gamma_\mathrm{b} = 103$.}
              \fig{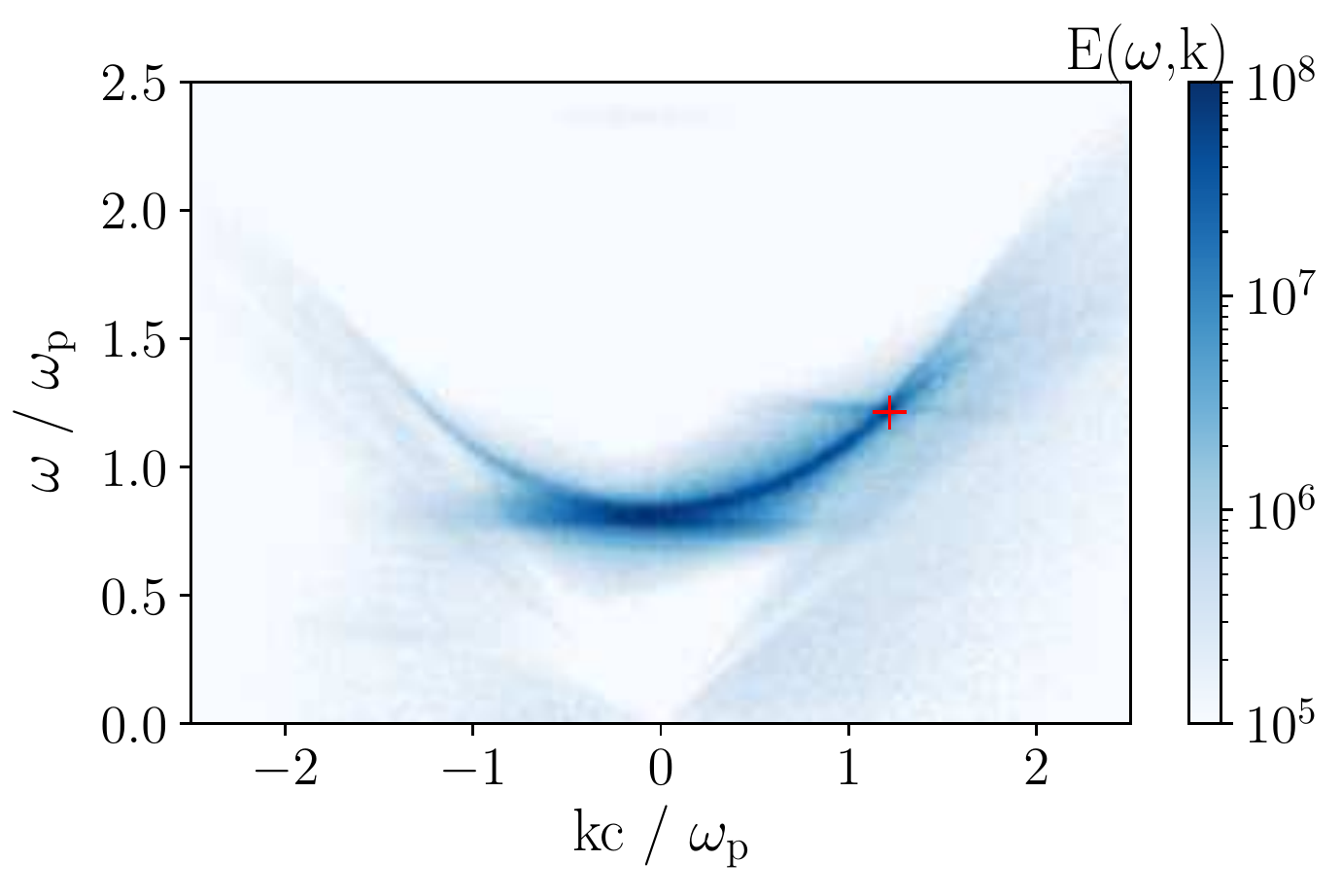}{0.333\textwidth}{(c) $\rho_0 = \rho_1 = 3.33, \gamma_\mathrm{b} = 60$.}
             }
    \caption{Dispersion diagrams of the electric field component.
    The red ``plus'' symbol denotes the position of the most unstable wave predicted by analytical theory.
    Same simulations as in Figure~\ref{fig1}.}
    \label{fig4}
\end{figure*}

Figure~\ref{fig4} shows dispersion diagrams of the electrostatic
waves for the whole simulation domain and for the entire simulation time.
We use every 20th time step to calculate the Fourier transform in the time-space domain.
Two main electrostatic branches of the longitudinal waves are present in the simulations:
the subluminal L-mode branch with s$\omega \approx k c$ (diagonal)
and the superluminal L-mode branch that has a (horizontal) cut-off frequency
$\omega \rightarrow \omega_0$ as $k \rightarrow 0$.
The superluminal branch bends towards higher frequencies and approaches
$\omega \approx k c$ as $k \rightarrow \infty$.
In Case~1, the waves are localized at $\omega / \omega_\mathrm{p} \approx k c / \omega_\mathrm{p} \sim 1.65$
along the subluminal dispersion branch. The electrostatic waves
at the superluminal branch are visible, i.e., their spectral power is slightly
enhanced with respect to the surroundings in the dispersion diagram
due to the thermal noise that excites waves along the plasma normal modes.
In Cases~2 and 3, the spectral power along the superluminal branch is more enhanced than for the Case 1.
Those waves have phase speeds larger than the speed of light.
Note that a new type of horizontal wave branch appears near the frequency $\omega_0$
and for the wave number range $k c / \omega_\mathrm{p} \sim (-1,1)$.
This mode corresponds to the formed solitons, it is a product of nonlinear effects. The effects of its intersection with the (undamped) light wave line and/or possibly with higher-order damped subluminal modes \citep{Godfrey1975a} cannot be easily predicted by analytical linear theories, since the distribution function already significantly deviates from the initial distribution at the soliton formation time.
Red pluses denote the frequency and wave number of the waves
that have the highest values of the analytically computed growth rate as in Figure~\ref{fig3}.
They show the point where the instability starts to develop
as predicted by linear theory.

\subsection{Soliton Properties}

\begin{figure}
    \centering
    \includegraphics[width=0.5\textwidth]{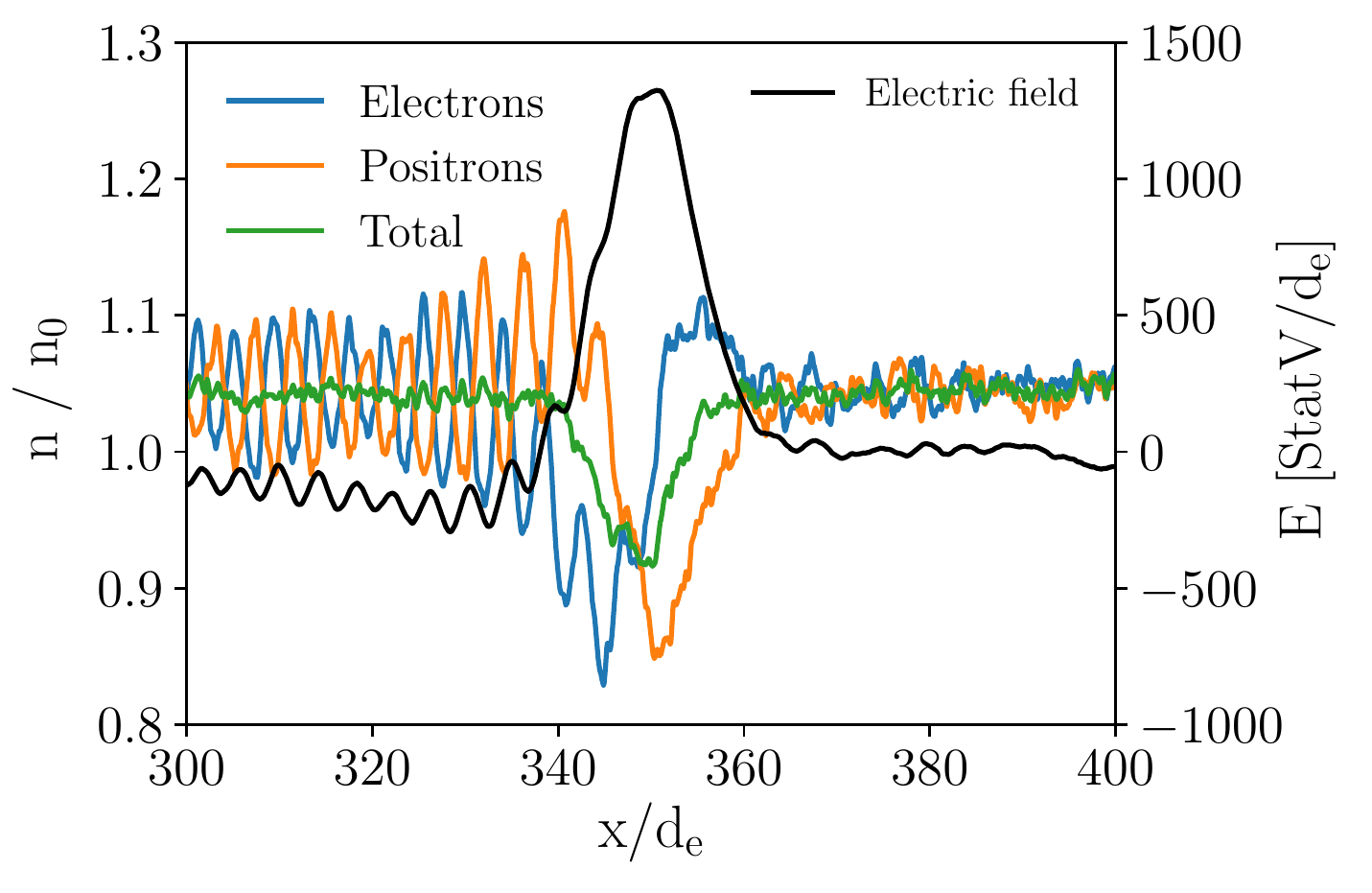} \\
    \includegraphics[width=0.5\textwidth]{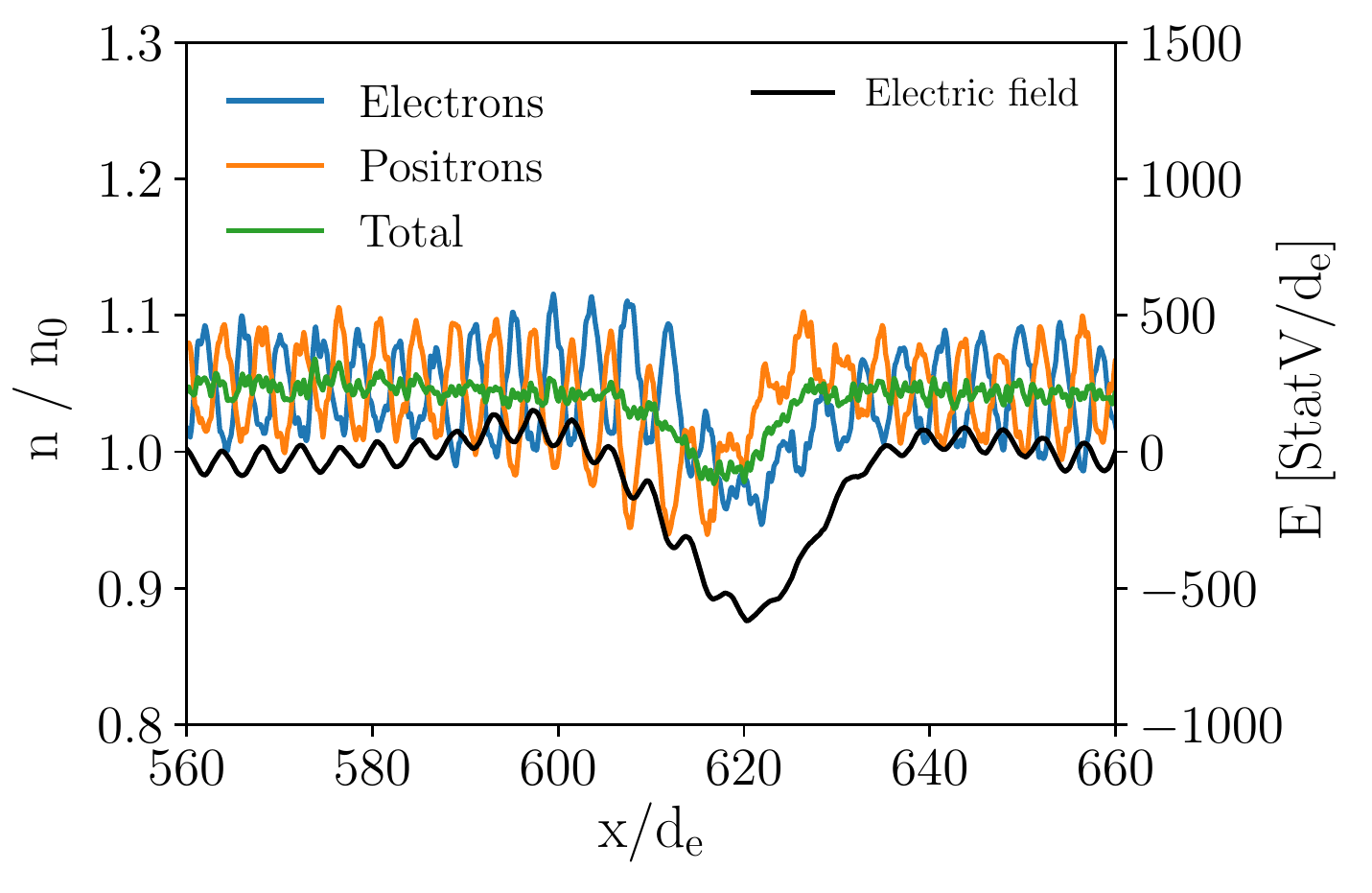}
    \caption{Spatial profiles of the electron, positron,
    total particle density, and the electric field for plasma parameters
    in Case~2 and two selected solitons in positions $x = 500-1800$
    (top) and $x = 4700 - 6000$ (bottom) at time $\omega_\mathrm{p} t = 14076$.
    Collate with Figures~\ref{fig6} and \ref{fig7}.}
    \label{fig5}
\end{figure}
In order to further investigate the soliton properties,
we selected two solitons at the time $\omega_\mathrm{p} t = 14076$ from the simulation Case~2.
Figure~\ref{fig5} shows the particle density profiles for the electron, positron, and total density.
Densities are normalized to the initial particle density of each species separately.
The first soliton is located at $x \sim 350\,d_\mathrm{e}$.
At this position, both species form a density depression.
In the left wing of the soliton ($x = 341-348\,d_\mathrm{e}$),
the positron density   increases while the electron density decreases.
In the right soliton wing ($x = 352-362\,d_\mathrm{e}$), the situation is the opposite.
To the left of the soliton, the electrons and positrons
form density waves with an amplitude of $\Delta n / n_0 \sim 0.05$.
The waves are mutually shifted in phase in such a way that the total density only has small fluctuations.
The situation is, however, different to the right of the soliton.
There are only small variations of the individual electron and positron densities. 
The second soliton is located at the position $x \sim 620\,d_\mathrm{e}$,
and it is also associated with a density depression.
The density of both species is nearly the same in both soliton wings.
Outside of all the formed solitons, both species form phase shifted
density waves with an amplitude of $\Delta n / n_0 \sim 0.05$.
The typical length of both solitons $l$, measured as full width at half maximum,
is nearly the same for both solitons, $l = 9.95\,d_\mathrm{e}$ ($k c / \omega_\mathrm{p} = 0.63$)
and $l = 10.31\,d_\mathrm{e}$ ($k c / \omega_\mathrm{p} = 0.61$), respectively.
For the first soliton, the electric field is positive at the density depression. 
The electric field to the left of this soliton is slightly negative, while
the particle charge density fluctuates.
On the other hand, the electric field is negative at the density depression of the second soliton.

\begin{figure*}
    \centering
    \includegraphics[width=0.7\textwidth]{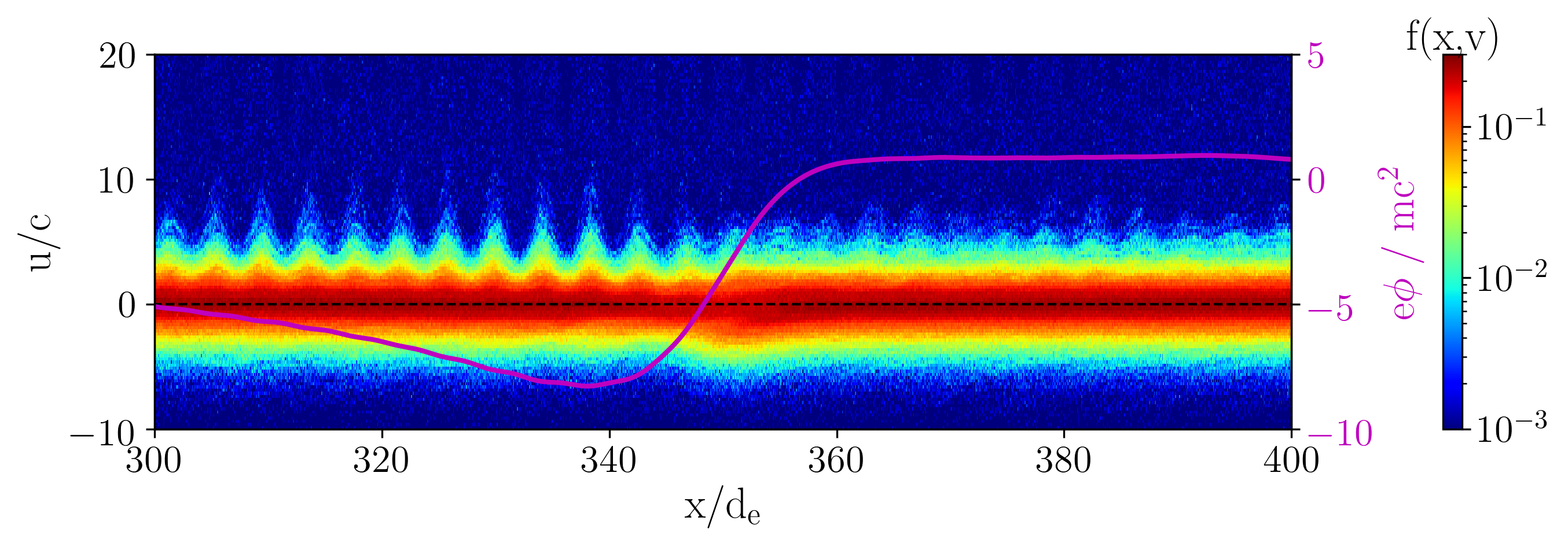} \\
    \includegraphics[width=0.7\textwidth]{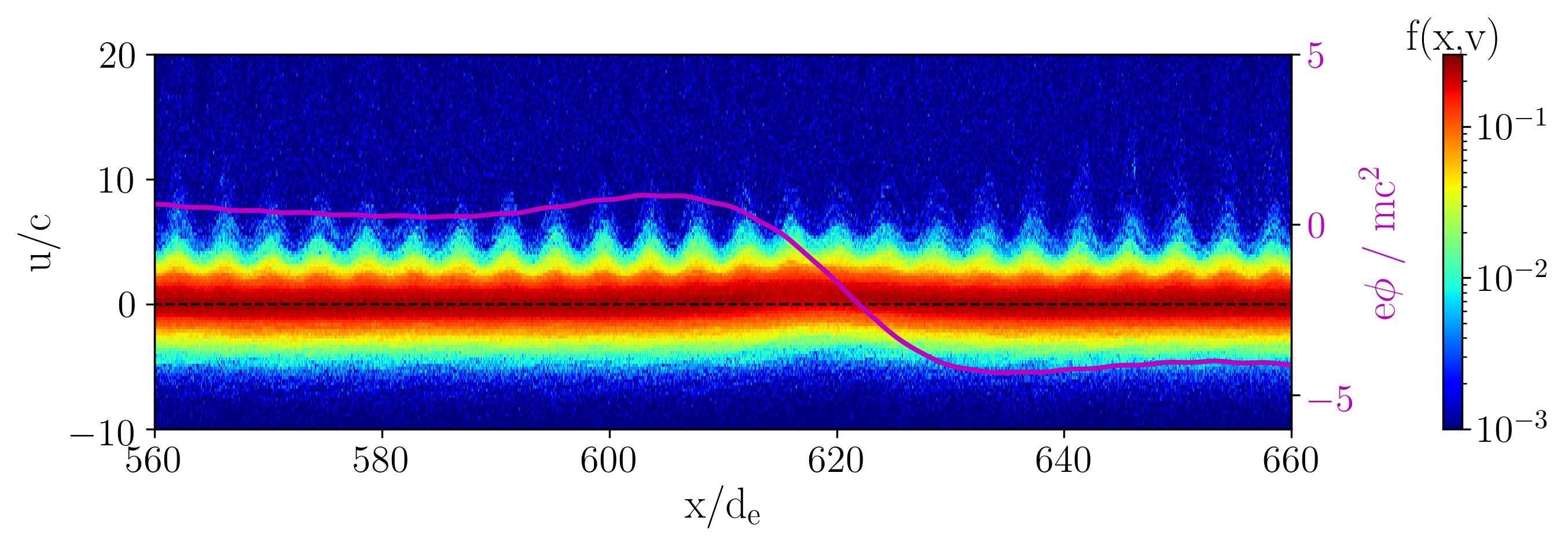}
    \caption{ 
    Electron phase space in \emph{logarithmic} scale for the two solitons
    from Figure~\ref{fig5} normalized to the initial electron density.
    \emph{Magenta line:} Electric potential normalized to particle rest mass energy.}
    \label{fig6}
\end{figure*}
Figure~\ref{fig6} shows the electron phase space for both
solitons normalized to the initial particle density.
The particles create spatial waves with positive velocities $u/c \sim 3-10$,
while for negative velocities, the distribution is more homogeneous in phase space. 
At the location of the first soliton ($x \sim 347-355\,d_\mathrm{e}$),
the background distribution spread is slightly broader indicating plasma heating.
Note the opposite direction of the electrostatic potential jump between those two solitons.
The height of the potential jump is $e\phi / mc^2 \approx 9.2$ and $e\phi / mc^2 \approx 5.2$, respectively.

\begin{figure*}
    \centering
    \includegraphics[width=0.7\textwidth]{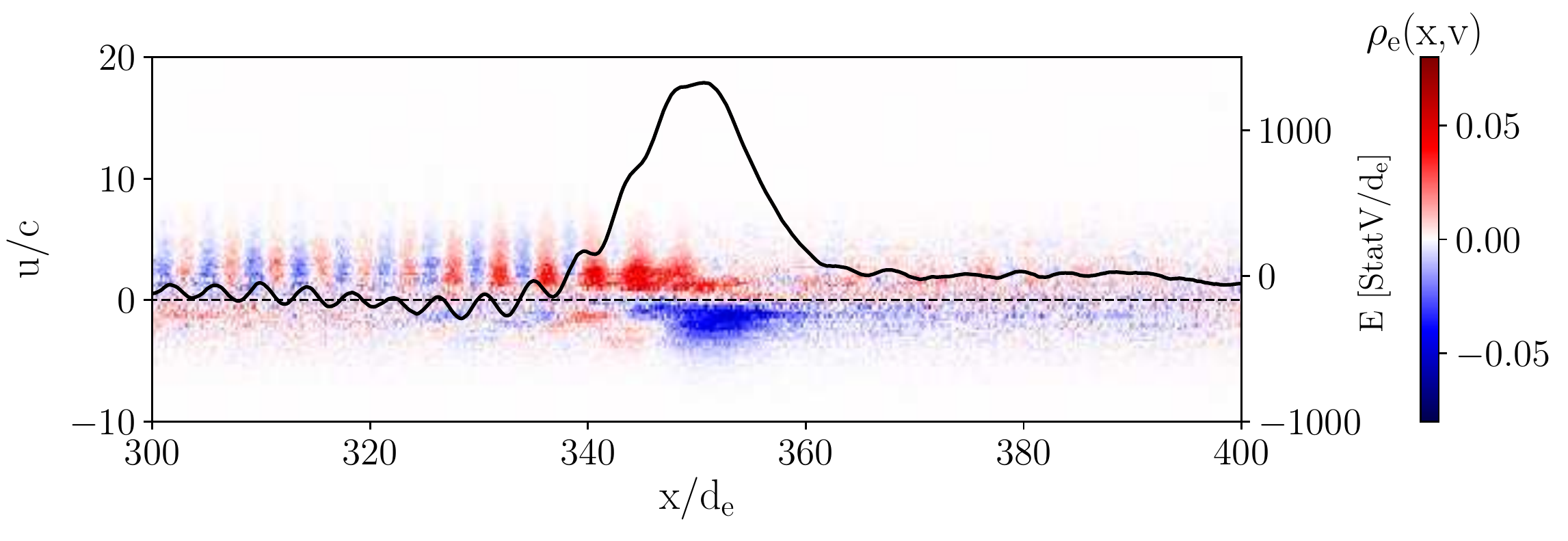} \\
    \includegraphics[width=0.7\textwidth]{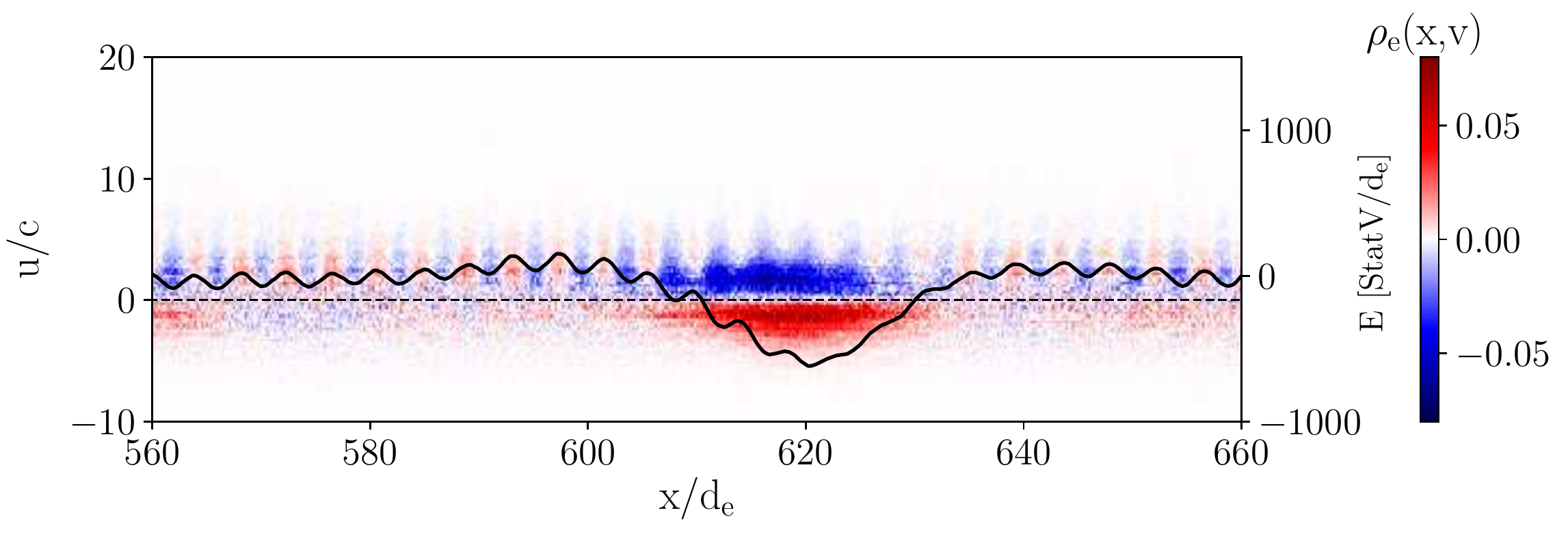}
    \caption{
    Phase space charge density of electrons and positrons
    for the same two solitons from Figure~\ref{fig5}.
    The charge density is normalized to the initial particle density and has a \emph{linear} scale.
    \emph{Red regions:} Phase space mostly occupied by positrons, \emph{Blue regions:}
    Phase space mostly occupied by electrons. \emph{Black line:} Electric field. }
    \label{fig7}
\end{figure*}
Figure~\ref{fig7} shows the charge density in the phase space.
The red color regions represent phase space locations with a positive electric charge,
i.e., more positrons than electrons.
In the blue regions, the opposite is the case, i.e., there are more electrons than positrons.
The charge density is normalized to the initial particle density.
In the first soliton, positrons have mainly positive velocities while electrons have negative ones.
Moreover, the electrons are located to the right of the electrons.
That creates a positive electric field.
For the second soliton, positrons have mostly negative velocities while electrons have mostly positive ones.
The electrons are located left from the positrons,
creating a weaker electric field in the negative direction. 
Outside of the solitons, for $u/c < 0$ both species are distributed relatively homogeneously,
while for $u/c > 0$ the charge density forms alternating clusters in space.
Oscillations are not fully visible for the first soliton for $x > 350\,d_\mathrm{e}$,
because the oscillating species are in such a relative phase that their total charge is almost zero.

\begin{figure}
    \centering
    \includegraphics[width=0.5\textwidth]{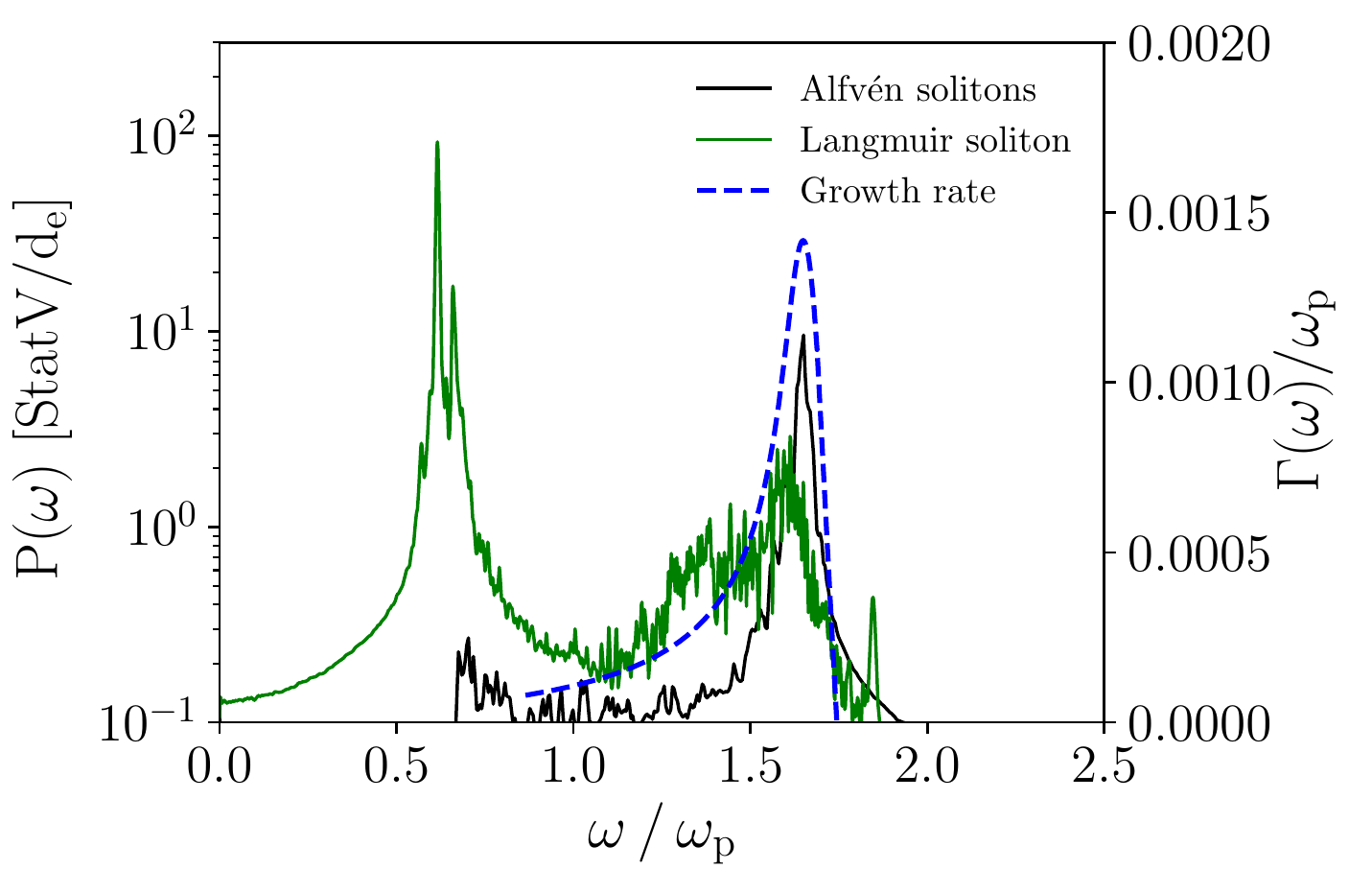}
    \caption{Spectral power of solitons that appear in simulations
    representing both soliton types as function of the frequency.
    Simulation parameters same as in Figure~\ref{fig5}.
    \emph{Black solid line:} Superluminal solitons during saturation Type~1 for the time interval
    $\omega_\mathrm{p} t = 3350 - 4690$ and the whole simulation box $x = 0 - 710.8$.
    \emph{Green solid line:} Saturation Type 2 case (only the second soliton is selected)
    for the time interval $\omega_\mathrm{p} t = 11400 - 15400$
    and position $x = 312.752 - 390.94$.
    \emph{Blue dashed line:} Analytically computed growth rates
    as a function of frequency with linear scale displayed on the right.}
    \label{fig8}
\end{figure}
The spectral power in solitons as a function of the frequency is shown in Figure~\ref{fig8}. 
We take all subluminal solitons that occur in saturation Type~1
and the selected superluminal soliton in saturation Type~2.
The spectral power densities are normalized as $P(\omega) = E(\omega)/ (L \delta t)$,
where $L$ is the spatial length and $\delta t$ the time interval that are used for the Fourier transform.
Thus the spectral power densities can be compared although
they were calculated with different time intervals and/or spatial sizes.
For saturation Type~1 the peak intensity is at the frequency $\omega / \omega_\mathrm{p} \sim 1.65$
in subluminal solitons and at frequency $\omega_0 \sim 0.616$ in superluminal solitons.
The blue dashed line denotes analytically computed growth rates.
It is expected that these waves already appear during the linear growth phase, before the formation of solitons.

\begin{figure}
    \centering
    \includegraphics[width=0.49\textwidth]{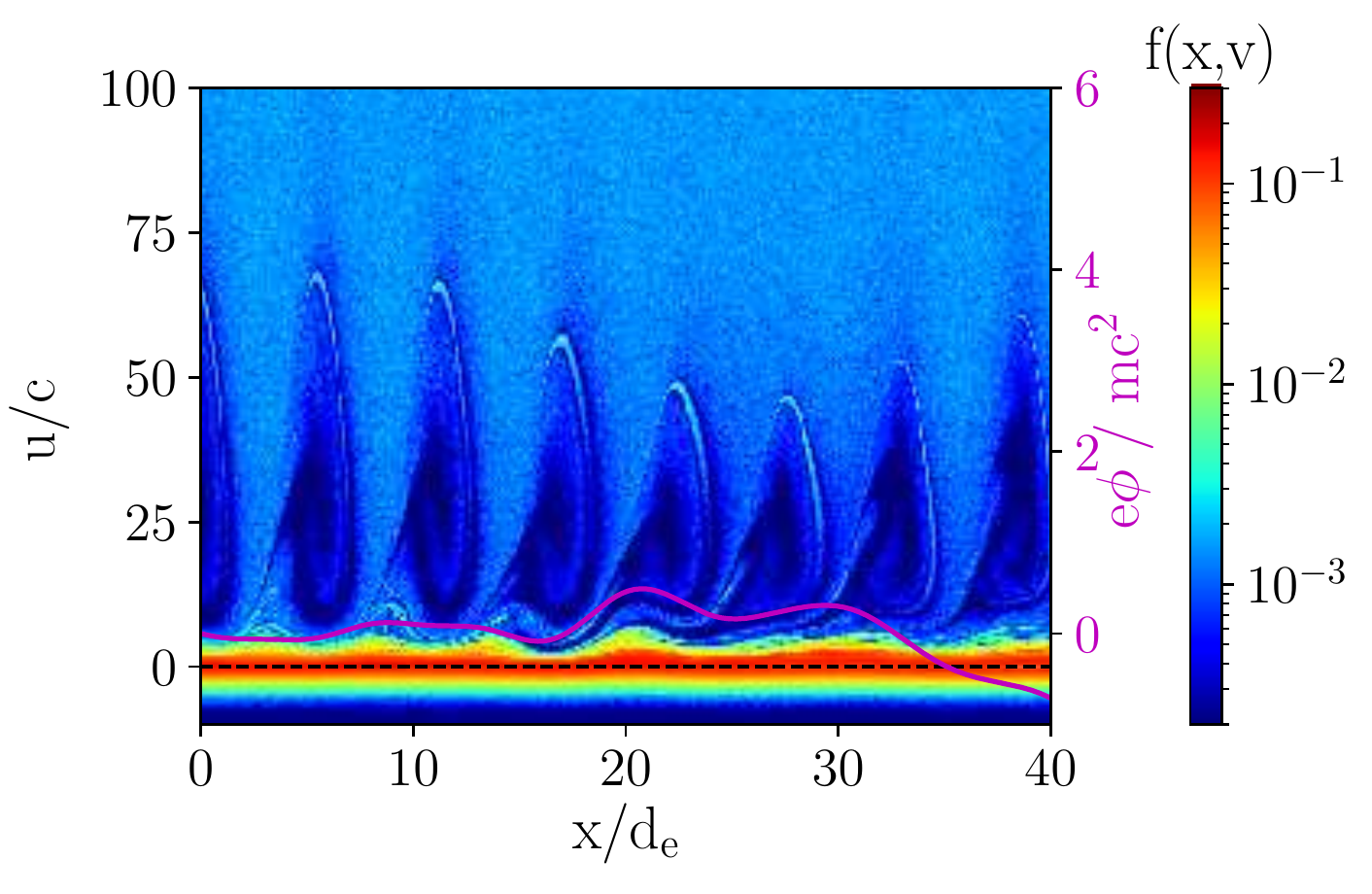}
    \caption{Electron phase space of subluminal solitons in a selected part of the simulation domain.
    This was obtained for a simulation with saturation Type~2 case
    for $n_0 = n_1$, $10^{5}$~electrons per cell, $\rho_0 = \rho_1 = 1, \gamma_\mathrm{b} = 103$.
    The particle density is normalized to the initial electron density. 
    \emph{Black dashed line:} $u/c = 0$.}
    \label{fig12}
\end{figure}
Figure~\ref{fig12} shows the electron phase space with an example of one subluminal soliton.
In order to highlight the generated phase holes, this figure
was obtained for another simulation run with $10^5$ electrons and $10^5$ positrons per cell,
and a density ratio of $n_0 = n_1$.
All other parameters remain the same, i.e., $\rho_0 = \rho_1 = 1, \gamma_\mathrm{b} = 103$.
The data to generate this figure were taken for the time $5630\,\omega_\mathrm{p}t$. 
We can see that the size of the electron holes along the simulation changes.
The electrons are typically oscillating in the velocity range $u/c = 5 - 70$.
The typical soliton sizes are $\delta x \sim 5\,d_\mathrm{e}$,
corresponding to $k c / \omega_\mathrm{p} = 1.26$. It is approximately
the wave number of the initial unstable waves that are predicted by analytical theory \citep{Manthei2021} as $k c / \omega_\mathrm{p} = 1.2166$.
Note that in this specific case, the simulation has a density ratio $n_0 = n_1$,
so that the wavelength of the initial unstable subluminal waves is
different from the studied Case~2 (see, e.g., Figure~\ref{fig3}b).

\subsection{Growth Rates and Saturation Energies}
In this subsection we analyze the growth rates and the saturation
electrostatic energies of simulations for a broader range of plasma parameters
than the previously shown examples.
\begin{figure}
    \centering
    \includegraphics[width=0.49\textwidth]{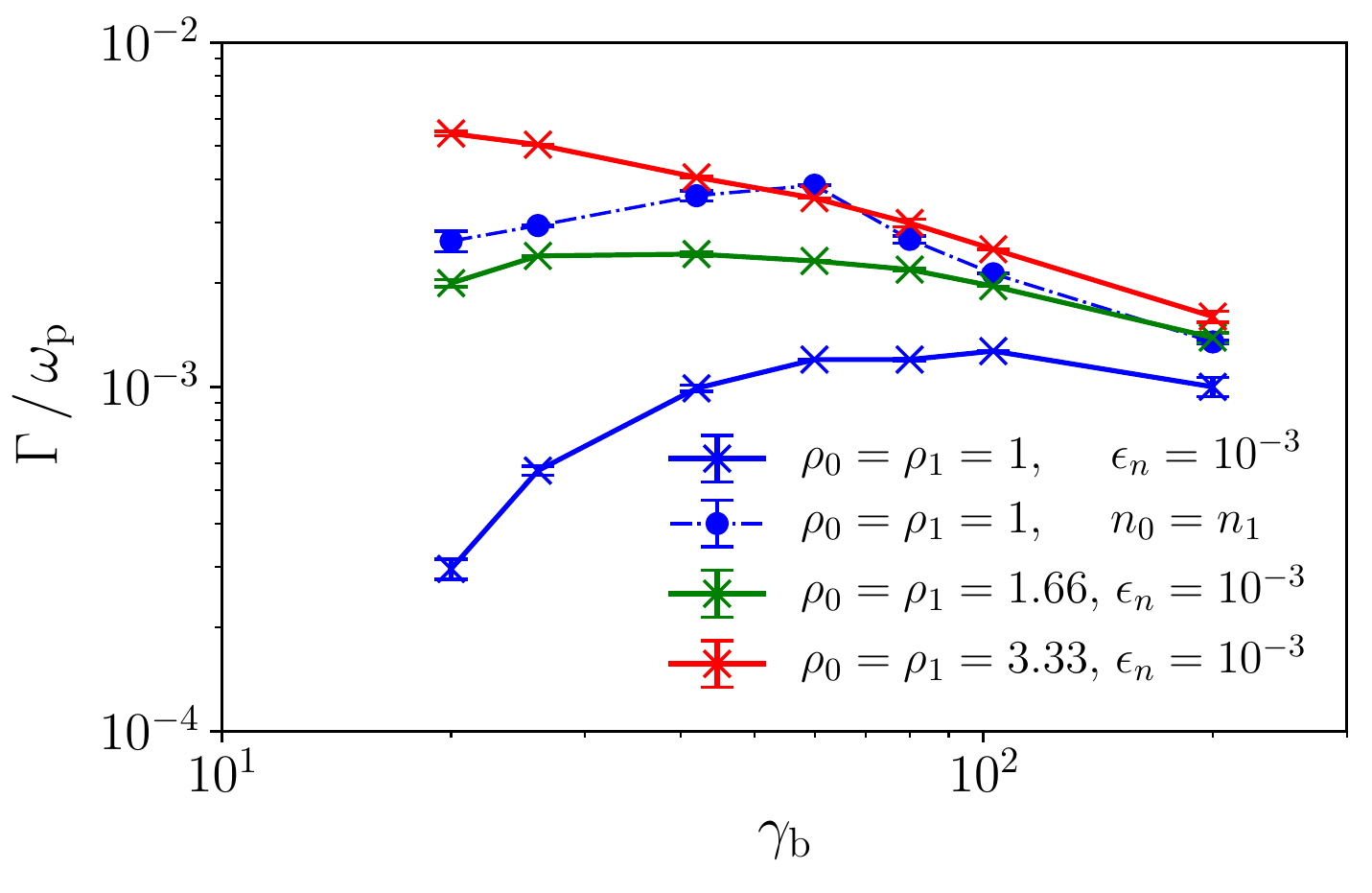}
    \caption{
    Integrated growth rates of the electrostatic waves as a function
    of the beam Lorentz factor for different plasma inverse temperatures.}
    \label{fig9}
\end{figure}

Figure~\ref{fig9} shows the growth rates of the initially growing electrostatic waves.
The growth rates $\Gamma$ are estimated from an exponential fit
to the electrostatic energy curves as a function of time.
We choose a fit function in the form $f(t) = E_\mathrm{0} + E_1 \mathrm{e}^{2 \Gamma t}$,
where $E_0$, $E_1$, and $\Gamma$ are fit parameters.
The growth rate uncertainties are estimated from the fitting procedure itself.
The growth rates are plotted as a function of the temperature
(solid lines with crosses) and as a function of the density (dash-dotted lined with circles)
for cases with $r_\mathrm{n} = 10^{-3}$ and $n_0 = n_1$.
The growth rates increase with decreasing temperature,
although only weakly for higher beam velocities.
With increasing beam velocity ($\gamma_\mathrm{b} \rightarrow \infty$),
the beam and the background distributions do not overlap anymore, and as a consequence,
the growth rates as a function of the beam velocity approach
an arbitrary common asymptote independent from their temperature.
\citet{Manthei2021} also found such an asymptotic behavior
of the growth rates as a function of the beam and background temperature
by means of linear theory and PIC simulations.
Note that the beam density ratio between the blue solid
and the blue dash-dotted lines decreases with the beam velocity
in the background reference frame.
For these lines, the density ratio is 50 and 5  for $\gamma_\mathrm{b} = 20$
and $\gamma_\mathrm{b}=200$, respectively.

\begin{figure*}
    \centering
    \gridline{
              \fig{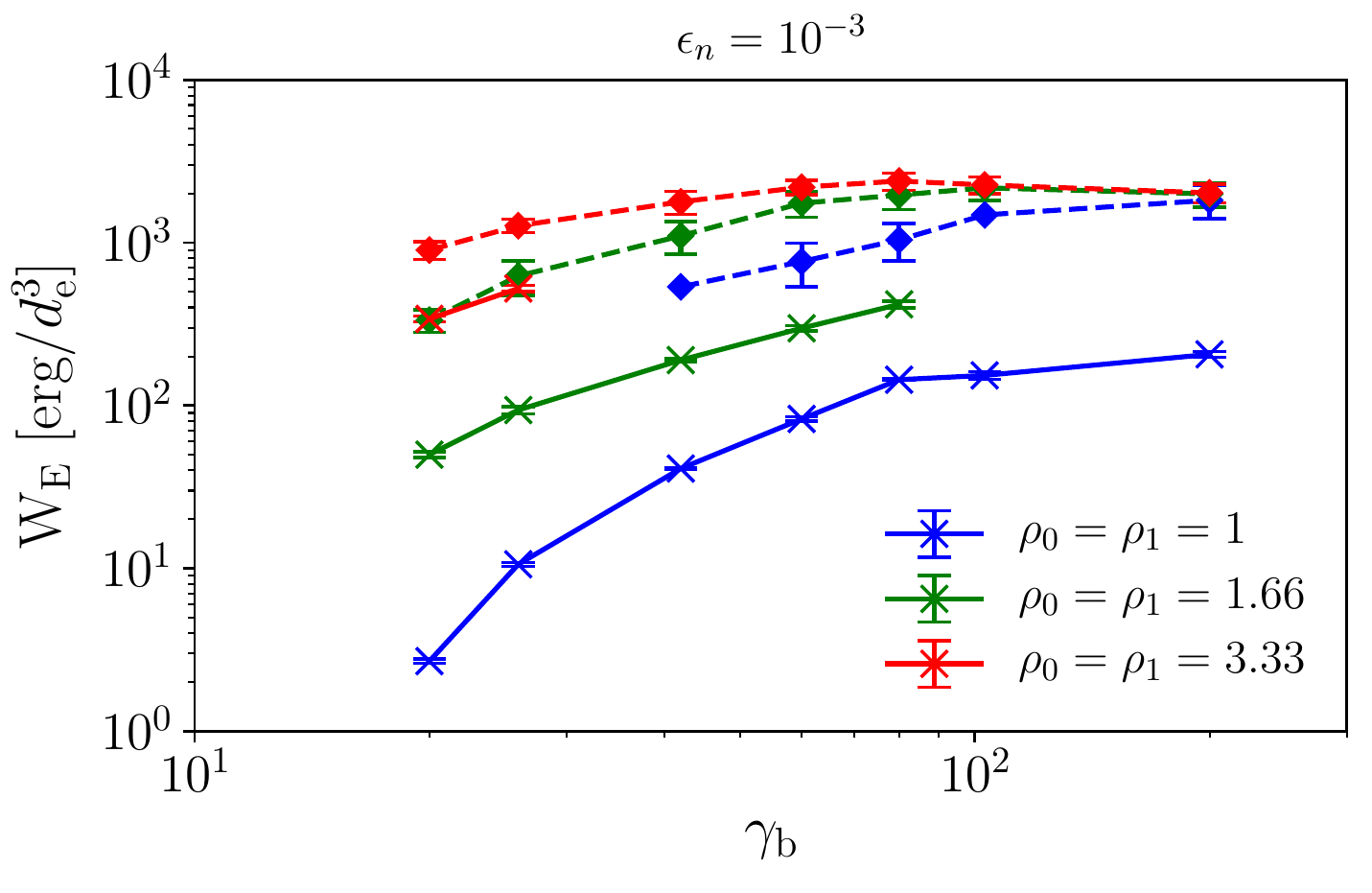}{0.49\textwidth}{(a) Constant $r_\mathrm{n} = 10^{-3}$.}
              \fig{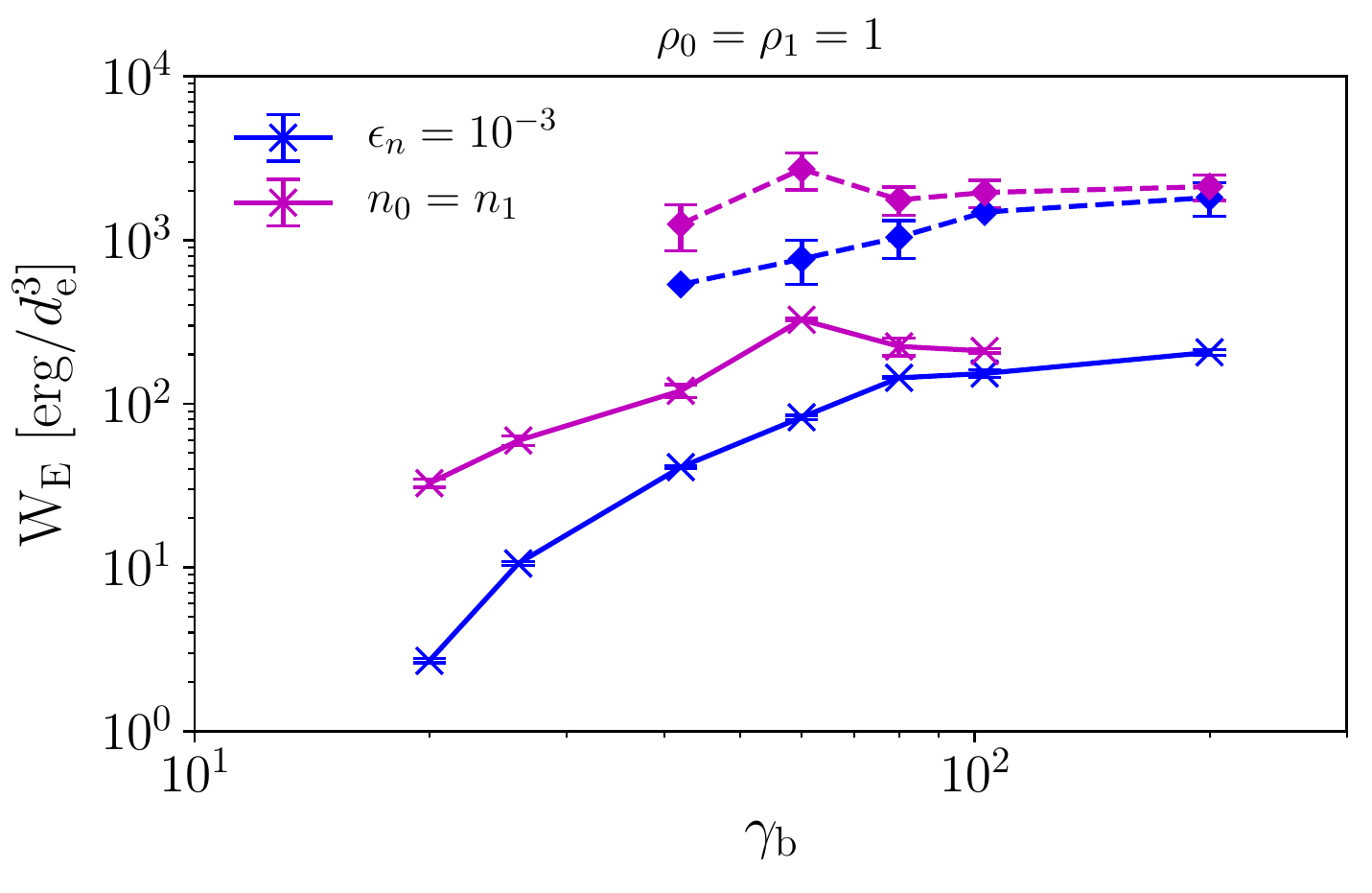}{0.49\textwidth}{(b) Constant $\rho_0 = \rho_1 = 1$.}
             }
    \caption{ Saturation energy density of the electrostatic waves
    for simulations with both saturation types (when applicable)
    as a function of the beam velocity.
    The energy density is normalized to the volume of a skin depth cube.
    \emph{Solid lines with crosses:} Saturation energies for Type 1.
    \emph{Dashed lines with diamonds:} Saturation energies for Type 2.
    The panels represent keeping separately constant $r_\mathrm{n} = 10^{-3}$
    (a) and $\rho_0 = \rho_1 = 1$ (b).}
    \label{fig10}
\end{figure*}
The electrostatic energy densities are shown in Figure~\ref{fig10}
as a function of the beam velocity for three plasma temperatures and two density ratios.
The energy density for the saturation Type~1 case is denoted by solid lines with crosses.
The saturation Type~2 case is represented by the dashed line with diamonds.

\begin{figure*}
    \centering
    \gridline{
              \fig{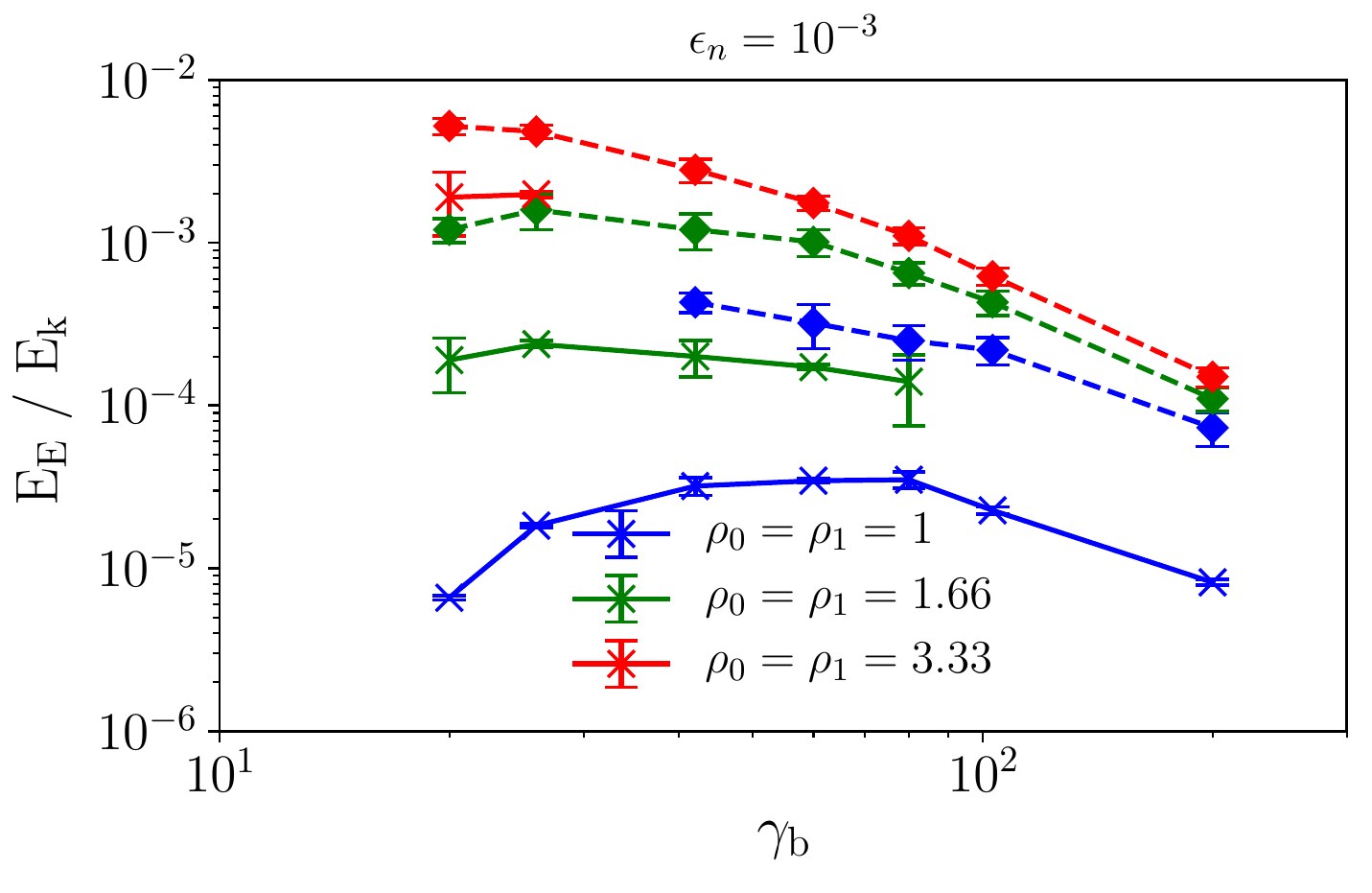}{0.49\textwidth}{(a) $r_\mathrm{n} = 10^{-3}$.}
              \fig{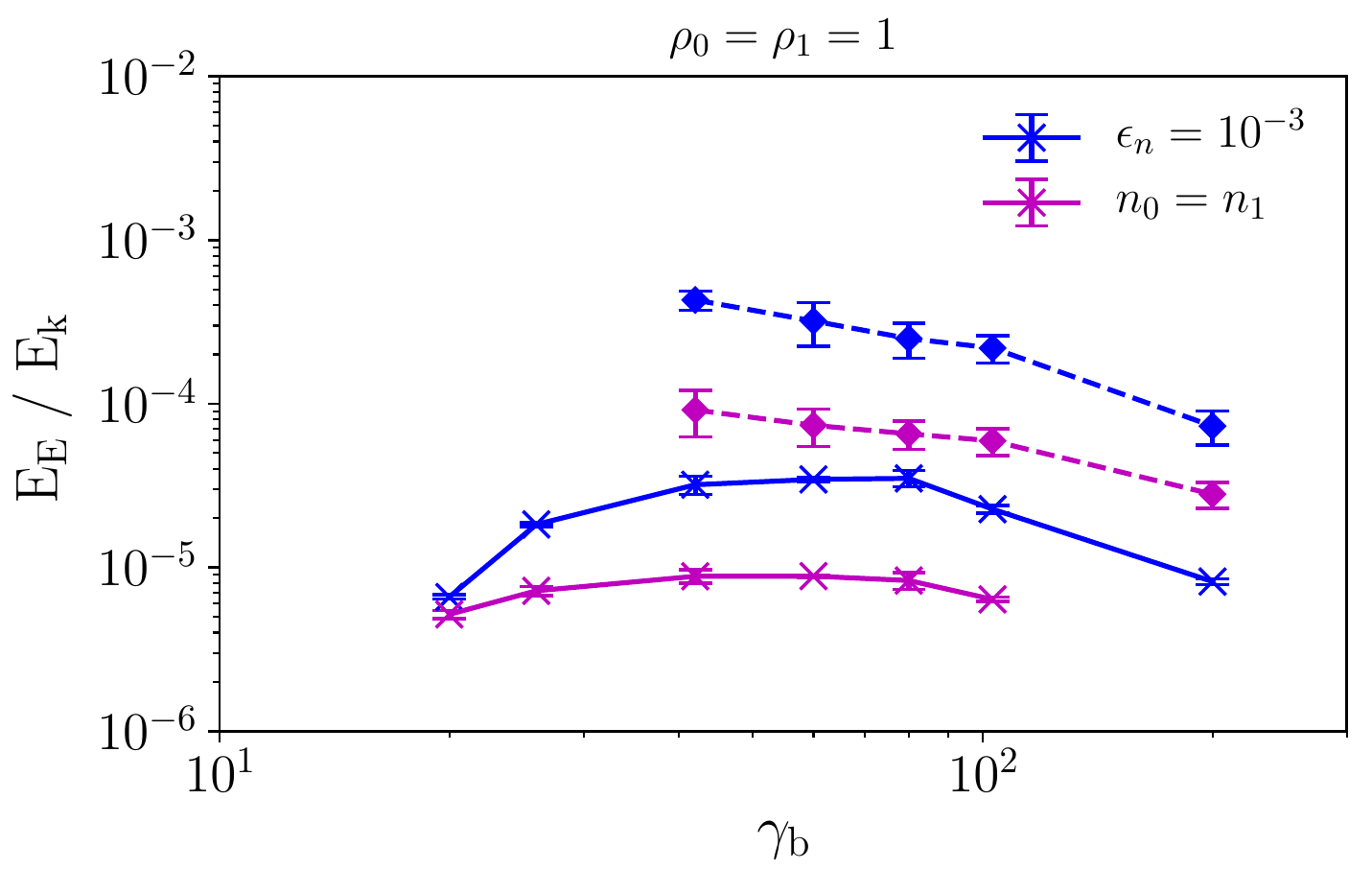}{0.49\textwidth}{(b) $\rho_0 = \rho_1 = 1$.}
             }
    \caption{Ratios of the electrostatic to the kinetic energy
    for both saturation types simulations as a function of the beam velocity for two density ratios.
    \emph{Solid lines with crosses:} Saturation energy ratios
    for saturation Type 1. \emph{Dashed lines with diamonds:}
    Saturation energy ratios for saturation Type 2.
    In each panel, a different quantity is kept constant:
    $r_\mathrm{n} = 10^{-3}$ in panel (a) and $\rho_0 = \rho_1 = 1$ in panel (b).}
    \label{fig11}
\end{figure*}
Figure~\ref{fig11} shows the electrostatic energy density normalized to the initial kinetic energy density.
The energy density ratio is, of course, a dimensionless number.
It shows how much of the initial kinetic energy is transformed
into electrostatic energy, and it is the same in all coordinate systems. 

\section{Discussion} \label{sec:discuss}

We used Particle-in-Cell simulations to study the relativistic streaming instability
in a relativistically hot pair plasma as a source for the pulsar radio emission.
After the initial exponential increase of the electrostatic energy,
the instability manifests two saturation types.
In plasmas with a high temperature background, $\rho < 1.66$,
the unstable subluminal wave saturates at low energy.
Only a small amount of the initial kinetic energy is released.
However, for the case of a colder plasma ($\rho \geq 1.66$,
but still relativistically  hot), or high beam velocities,
a different saturation occurs, that we denoted as Type~2.
This type of saturation has a much higher electrostatic energy density than the previous one.
It leads to the formation of large-amplitude L-mode waves and solitons.
During the evolution, the initially unstable subluminal waves quickly
spread up into the superluminal part of the L dispersion branch
with most of the electrostatic energy localized in the range $k c / \omega_\mathrm{p} = 0 - 0.5$.
The frequency and wave number of the initially most
unstable wave is in a very good agreement with linear theory.
We estimated the wave numbers: $k_1c / \omega_\mathrm{p} \sim 1.65$,
$k_2c / \omega_\mathrm{p} \sim 1.65$, and $k_3c / \omega_\mathrm{p} \sim 1.20 $ for Cases~1--3.
The linear theory predicts values  $k_1c / \omega_\mathrm{p} = 1.6649$,
$k_2c / \omega_\mathrm{p} = 1.6510$, and $k_3c / \omega_\mathrm{p} = 1.2172$ \citep{Manthei2021}, see Figures~\ref{fig3} and~\ref{fig4}.
Although the linear calculations can predict the position
of the initially most unstable waves well,
they are not applicable to predict the formation
and nonlinear properties of the superluminal L-mode waves (Figure~\ref{fig8}),
since the exponential growth of the wave becomes eventually
too large to be considered as of small amplitude, an assumption
made when performing the linearization. 
Moreover, the formation of density depressions connected with the initially unstable waves plays an essential role for the soliton formation.
During the saturation stage,
the system is rather dominated by non-linear effects.

Moreover, we found that stable superluminal large amplitude L-mode waves
are excited only during the second saturation type. 
The size of the solitons, estimated as the full width at half maximum,
increases with increasing plasma temperature.
Also, the energy of its accompanying electric field increases
with increasing beam velocity. Solitons mutually interact;
they merge and can separate. If they are distant enough from other solitons,
they can be stable from their creation time on until the end of the simulation,
i.e., $>1000\,\omega_\mathrm{p}t \gg \tau_\mathrm{p}$,
where $\tau_\mathrm{p}$ is the plasma period, which is a necessary condition for the soliton stability. 

While the subluminal solitons are associated with electron and positron density oscillations,
which can be mutually shifted in space,
while keeping the total density approximately constant,
the superluminal solitons are localized in the density depressions of both species.
They are associated with a strong oscillating electric field and charge bouncing.
The associated electrostatic energy can be more than one order
of magnitude higher than in the surrounding plasma.
We verified that the value 
of this energy does not depend significantly on the number of particles per cell.
The typical soliton size is, e.g., $\sim 10.3\,d_\mathrm{e}$
for $\rho_0 = \rho_1 = 1, \gamma_\mathrm{b} = 103$ ($\sim 31$~cm for $\omega_\mathrm{p} = 10^{10}$).  
For a plasma colder than in this example, the L-mode soliton size becomes smaller. 
\citet{Weatherall1997} estimated the half width at half maximum size (HWHM) as $3.5\,d_\mathrm{e}$;
however, all their solitons have a positive group velocity $\sim 0.071~c$.
In our simulations, superluminal solitons have small positive and negative
group velocities in comparison with the speed of light.
The soliton  sizes are also in agreement with \cite{Melikidze2000},
who used a different approach for soliton formation and estimated its size as $\sim10-100$~cm. 
On the other hand, from a simple radio propagation model,
the estimated size of the emission region is 12~cm for 0.4~ns pulsar nanoshots \citep{Hankins2007}. 
Therefore, these nanoshot emission regions could be associated with solitons.

\begin{deluxetable*}{l|llc|c|ccc}
\tablecaption{Computed electrostatic energy density
for three plasma frequencies in the pulsar reference frame with Lorentz factor $\gamma$.
Three main cases are selected: (1) Mean electrostatic energy density
in the simulation corresponding to the energy density in subluminal solitons.
(2) Electrostatic energy density in the superluminal soliton only.
(3) Ultrarelativistic superluminal soliton with Lorentz factor $\gamma=10^6$ (corresponding to the primary beam).}
\label{tab1}
\tablewidth{0pt}
\tablehead{
Type & \colhead{$\gamma_\mathrm{b}$} & \colhead{$\rho_0,\rho_1$} & $E_\mathrm{E} / E_\mathrm{k}$ & $\gamma$ &
\multicolumn3c{Electrostatic energy density [erg$\cdot$cm$^{-3}$]} \\
 &  &  &  &  & \colhead{$\omega_\mathrm{p} = 3.6\times10^9$} & \colhead{$\omega_\mathrm{p} = 10^{10}$} & \colhead{$\omega_\mathrm{p} = 7.9\times 10^{11}$}
}
\startdata
Mean electrostatic & $26$ & $1$ & $1.7\times 10^{-5}$ & $10^2$ & $5.6\times 10^0$ & $4.3\times 10^1$ & $2.7\times 10^5$ \\
energy & $103$ & $1$ & $2.5\times 10^{-4}$ & $10^2$ &  $8.1\times 10^1$ & $6.3\times 10^2$ & $3.9\times 10^6$ \\
OR & $60$ & $3.33$ & $1.8\times 10^{-3}$ & $10^2$ & $5.7\times 10^2$ & $4.4\times 10^3$ & $2.8\times 10^7$ \\
Subluminal soliton & $26$ & $3.33$ & $4.8\times 10^{-3}$ & $10^2$ & $1.6\times 10^3$  & $1.2\times 10^4$ & $7.5\times 10^7$ \\
\hline
Superluminal soliton & $103$ & $1$ & $4.2\times 10^{-3}$ & $10^2$ & $1.4\times 10^3$ & $1.1\times 10^4$ & $6.6\times 10^7$ \\
& $60$ & $3.33$ & $8.4\times 10^{-3}$ & $10^2$ & $2.7\times 10^3$ & $2.1\times 10^4$ & $1.3\times 10^8$ \\
& $26$ & $3.33$ & $4.5\times 10^{-2}$ & $10^2$ & $1.5\times 10^4$  & $1.1\times 10^5$ & $7.1\times 10^8$ \\
\hline
Primary beam &  &  &  &  &  &  &  \\
superluminal soliton & $26$ & $3.33$ & $4.5\times10^{-2}$ & $10^6$ & $1.5\times 10^8$  & $1.1\times 10^9$ & $7.1\times 10^{12}$ \\
\enddata
\end{deluxetable*}

We estimated the amount of electrostatic energy generated by the instability normalizing
in two different ways.
First, as energy in units of ergs divided by a volume that is equal
to a cube with its edge length equal to one skin depth.
This amount of energy is given in the simulation or background reference frame.
The second way is the normalization to the initial kinetic energy.
In the case of the subluminal solitons, up to 0.55~\%
of initial kinetic energy is transformed into electrostatic waves,
while for superluminal solitons, conversion efficiency amounts up to 4.5~\%.
Note that this ratio between the electrostatic and the kinetic energy
is a dimensionless quantity, and it is the same in all reference frames.
If we assume that the bunch is moving with a velocity corresponding to $\gamma = 100$
in the pulsar reference frame, the mean particle energy is $\sim 48.3$~MeV.
For the estimation of the total kinetic energy density,
we also need the number density in the pulsar emission region.
We assume three cases with densities corresponding to the plasma
frequencies $3.6\times10^9$~s$^{-1}$, $10^{10}$~s$^{-1}$,
and $7.9\times 10^{11}$~s$^{-1}$.
The first plasma frequency was considered by \citet{Usov1987},
the second one is for the typical pulsar magnetosphere
at the height of fifty neutron star radii,
and the third one is the plasma frequency in the magnetosphere
near the surface of the neutron star \citet{Weatherall1994,Eilek2016,Melrose2020a}.
For these values and four selected simulations,
we estimate the electrostatic energy density in the pulsar reference frame
and in the superluminal soliton (only three of these four simulations apply),
summarized in Table~\ref{tab1}.
For the typical pulsar plasma with frequency $10^{10}$~s$^{-1}$,
the energy densities are in the range $10^1 - 10^4$~erg$\cdot$cm$^{-3}$
for the subluminal solitons, and $10^4 - 10^5$~erg$\cdot$cm$^{-3}$ for the superluminal solitons.
If we take the most extreme case where the soliton is localized near the stellar surface
and $\gamma = 10^6$ (typical primary beam velocity),
the energy density can reach $7 \times 10^{12}$~erg$\cdot$cm$^{-3}$.
All cases are for $r_\mathrm{n} = 10^{-3}$ (weak beam).
Also, for higher $r_\mathrm{n}$ (strong beam regime),
the amount of generated energy increases, as can be seen from Figure~\ref{fig11}b.

Only a part of the electrostatic energy density can be converted into escaping electromagnetic waves.
Therefore, the generated electrostatic energy density should
be higher than the energy density of electromagnetic waves.
\citet{Weatherall1998} estimated a typical value of $0.2$~erg$\cdot$cm$^{-3}$
for the electromagnetic energy density of a typical pulsar pulse.
This value can be well explained by the radiation from the electrostatic
wave conversion in the subluminal solitons.
\citet{Jessner2005} estimated an energy density of $6.7 \times 10^{4}$~erg$\cdot$cm$^{-3}$
for the microsecond Crab giant pulses.
The microsecond pulse emission could be associated with superluminal and/or subluminal
solitons that carry electrostatic energy in the range $10^{4} - 10^{5}$~erg$\cdot$cm$^{-3}$.
For the necessary energy density of the nanosecond pulses  estimated from observations,
$ > (10^{12} - 10^{15})$~erg$\cdot$cm$^{-3}$ \citep{Soglasnov2004}
and $2\times 10^{14}$~erg$\cdot$cm$^{-3}$ \citep{Hankins2007},
it is also necessary to assume a very high density plasma near
the stellar surface as well as the primary beam velocity as the cause of the radio emission.

In a superluminal soliton, particles undergo strong acceleration in both directions
due to the strong electric field with amplitudes exceeding in some simulations
$10^3$~Statvolts ($3 \times 10^{5}$~V) per skin depth.
The behavior of particles in solitons is characterized by charge separation
in response to a strong electric field (Figures~\ref{fig5}--\ref{fig7}).
Due to the equal masses of both species in the electron-positron pair plasma,
the oscillations of both species are symmetric.
This process differs from the behavior of particles in ion-electron plasma solitons,
where mainly electrons oscillate in the presence of an ion density depression.
The typical velocity oscillations range from $u/c = -0.73$ $(\gamma = 1.24)$ up to $u/c = 0.63$ $(\gamma = 1.18)$ (Figure~\ref{fig7}). These oscillating particles can play a role in the pulsar linear acceleration emission \citep{Cocke1973,Fung2004,Melrose2009b,Reville2010}.
Those electron-positron relativistic L-mode solitons differ
from those found in the non-relativistic cold plasma approximation in several points.
First, our generated solitons have superluminal velocities.
The associated electric field is much stronger in our case.
While the ratio of electric to kinetic energy is usually about $10^5 - 10^{6}$ in near Earth space plasmas,
it can reach up to $0.045$ for our relativistic simulations with $r_\mathrm{n} = 10^{-3}$.
For non-relativistic temperatures, the typical soliton oscillation frequency is $\geq \omega_\mathrm{p}$
while in our relativistic solitons most of the energy is concentrated at frequencies $< \omega_\mathrm{p}$. 
In addition, the background-to-beam density ratio equal
to one is very often used in plasma simulations.
Nevertheless, we show that solitons are produced even in the weak beam approximation.

From our estimated soliton properties resulting from our simulations,
two electromagnetic emission processes can be considered.
In the relativistic plasma emission, the longitudinal electrostatic
subluminal waves ($\omega \approx k c$) and superluminal waves (close to $\omega_0 \equiv \omega(k = 0)$)
can interact with an external perpendicular modulation wave $S$
into electromagnetic transverse waves  $T$: $A + S \rightarrow T, L + S \rightarrow T$. 
\citet{Weatherall1997,Weatherall1998} estimated the timescale $\delta t$
of soliton ``collapse'' (formation) 
$\delta t \sim 1 / \gamma_s^{1/2}\Lambda \nu_p$ = $1 / \nu_{obs} \Lambda$,
where $\gamma_s$ is the Lorentz factor of a bunch moving in the pulsar reference frame,
$\nu_\mathrm{obs} = 2 \gamma_\mathrm{s}^{1/2} \nu_\mathrm{p}$ is the observed frequency,
and $\nu_\mathrm{p}$ is the plasma frequency in the bunch reference frame,
$\Lambda = E^2 / 8 \pi n_\mathrm{e} m_\mathrm{e} v_\parallel^2$
is the ratio between electrostatic to kinetic energy density
for the electric field amplitude $E$ and the electron density $n_\mathrm{e}$.
In our case $\Lambda \leq 0.042$ and the observed time scale $\delta t \geq 14.5$~ns
($\nu_\mathrm{p} = 10^{10}/2\pi$~s$^{-1}$).
This time is sufficiently small to form solitons in the time interval
of overlapping bunches $ \sim 10\,\mu$s or during the simulation time.
The typical frequency bandwidth is $\delta \nu / \nu_\mathrm{p} \approx \Lambda \leq 0.042$. 

The second possibility is the linear acceleration emission (LAE) by the particles oscillating in the solitons.
The electromagnetic waves are emitted as O- wave modes at the frequency $\omega \gamma^2$,
where $\omega$ is the oscillation frequency, and $\gamma$
is the Lorentz factor of oscillating particles, both in the simulation reference frame.
The emission is constrained to angles $\alpha \approx 1 / \gamma$ in the pulsar reference frame.
In our case, two possible emission sources apply: the superluminal solitons and subluminal solitons.
In the superluminal soliton from the simulation shown in Figures~\ref{fig6}--\ref{fig8},
the Lorentz factor varies between $\gamma = 1.24$ for negative velocities, and to $\gamma = 1.18$ for positive ones.
We expect that $\sim 90\%$ of particles undergo such oscillation.
In subluminal solitons, particles oscillate between $\gamma = 5$ and $\gamma = 65$,
but the particle density is $\ll n_0$, i.e., much less than in the case of superluminal solitons.

%
%
%

\section{Conclusions}
\label{sec:conc}

We studied the formation of solitary waves due to streaming
instabilities of relativistically hot electron-positron pulsar pair
plasmas as a possible source for the pulsar radio emission.
For this sake, we utilized Particle-in-Cell kinetic numerical
plasma simulations initializing one-dimensional
Maxwell-J\"uttner velocity space distributions for
appropriate pulsar plasma parameters.
Investigating the parameter range of inverse temperatures
$3.33 \leq \rho \leq 1$ and beam Lorentz factors
$20 \leq \gamma_\mathrm{b} \leq 200$, we found that
streaming instabilities develop, which, in the course of
their non-linear evolution, form for sufficiently high 
inverse temperatures $\rho \geq 1.66$ and 
sufficiently large relativistic $\gamma_\mathrm{b}$
stable large amplitude superluminal L-mode 
solitons, i.e. relativistically generalized Langmuir 
solitons.
The generated superluminal solitons are 
associated with density depressions and strong 
oscillations below the plasma frequency.
For the Type-2 saturation the electrostatic 
energy density of the waves reaches up to 
$1.1 \times 10^5$~erg$\cdot$cm$^{-3}$
for real pulsar parameters.
Note that superluminal solitons form 
only in relativistically hot plasmas at sufficiently 
low temperatures. Extremely hot plasmas are 
not expected to cause the observed intense 
pulsar radio emission by the formation of 
solitons.

Subluminal L-mode solitons are formed 
in all cases with wave numbers and frequencies
exceeding the plasma frequency, which were
theoretically predicted by~\citet{Rafat2019b}
and~\citet{Manthei2021}.
For the Type-1  saturation the energy density 
of the (electrostatic) subluminal solitons 
reaches, however,  only up to 
$1.2 \times 10^4$~erg$\cdot$cm$^{-3}$.

We expect that a large part of the electrostatic wave 
energy can be transformed into electromagnetic 
waves --- either by relativistic plasma emissions 
or the ``linear acceleration'' mechanism , i.e., 
either via wave-wave interactions or due to 
the acceleration of the particles in the strong 
electric fields of the solitary waves.

We conclude that the formation of superluminal
solitons can indeed be considered as a relevant 
mechanism of pulsar nanoshots and microsecond 
bursts.
The properties of the electromagnetic emission of such 
solitary waves should, however, be further investigated.

~
\acknowledgments
~

\vspace{-0.6cm}
\noindent We gratefully acknowledge the developers of the ACRONYM code
(\textit{Verein zur F\"orderung kinetischer Plasmasimulationen e.V.}),
and the financial support by the German Science Foundation (DFG) via 
projects MU-4255/1-1 and BU-777-17-1 as well as by
the Czech Science Foundation (GACR) via the project 20-09922J.
This work was supported by The Ministry of Education, Youth and
Sports from the Large Infrastructures for Research, Experimental
Development and Innovations project ``e-Infrastructure CZ – LM2018140''.
Part of the simulations were carried out on the HPC-Cluster of the
Institute for Mathematics of the TU Berlin.
Computational resources were also supplied by the project
"e-Infrastruktura CZ" (e-INFRA LM2018140) provided within the
program Projects of Large Research, Development and Innovations
Infrastructures.
The authors gratefully acknowledge the Gauss Centre for
Supercomputing e.V. (\url{www.gauss-centre.eu}) for partially funding
this project by providing computing time on the GCS Supercomputer
SuperMUC-NG at the Leibniz Supercomputing Center (\url{www.lrz.de}),
project pr27ta.
We also thankfully acknowledge the referee whose constructive 
comments and suggestions helped us to improve the presentation of 
our results.

\software{Python, PIC-code ACRONYM}



\appendix
\section{L-mode Waves} \label{AppendixA}

In order to predict the instability evolution analytically, we take advantage of the fact that a linearized version of the Vlasov-Maxwell system of equations \citep{melrose1986} provides a good approximation for the description of the initial, linear stage of the instability development. During this period, the wave amplitude is still sufficiently small such that higher-order effects can be neglected. Therefore, the linear theory discussed below only applies until the nonlinear affects prevails (i.e., before instability saturation). In the latter phase, the wave amplitude stops to grow. Such a system can only be studied via the conducted simulations while the following linear theory calculations contribute to the understanding of the instability growing phase.

The wave properties in a plasma can be described using the wave equation 
\citep{Melrose1999,Melrose1999b}
\begin{equation}
 \Lambda_{ij}(\omega,\boldsymbol{k}) e_j = 0,
\end{equation}
where $\Lambda_{ij}$ is the dielectric tensor
\begin{equation} \label{eq-A-1}
 \Lambda_{ij}(\omega,\boldsymbol{k}) = \frac{c^2 (k_ik_j - k^2 \delta_{ij})}{\omega^2} + K_{ij}(\omega, \boldsymbol{k}),
\end{equation}
and $\boldsymbol{e}$ is the polarization vector, $\omega$ is the wave frequency, $\boldsymbol{k}$ is the wave vector, $\delta_{ij}$ is the Kronecker delta. Equation \ref{eq-A-1} is solved by setting the determinant $|\Lambda_{ij}| = 0$.
Assuming a pulsar pair plasma, the wave vector in the $x-z$ plane, the waves in the low frequency limit $\omega \ll \omega_\mathrm{ce}$ (where $\omega_\mathrm{ce}$ is the electron cyclotron frequency), and neglecting non-gyrotropic terms, the determinant of the dielectric tensor, i.e., the dispersion relation, becomes
\begin{equation}
 |\Lambda_{ij}| = \Lambda_{22}(\Lambda_{11}\Lambda_{33} - \Lambda_{13}^2) = 0.
\end{equation}
The strength of the pulsar magnetic field allows to consider only parallel propagation $\boldsymbol{k} \parallel \boldsymbol{B}$, where $\boldsymbol{B}$ is the the magnetic field vector, such that the components $\Lambda_{13}$ and $\Lambda_{31}$ vanish due to their dependence on the propagation angle. Thereby,the dispersion relation factorizes into three independent equations, meaning that the remaining non-zero components $\Lambda_{11}$, $\Lambda_{22}$, and $\Lambda_{33}$ can be set to zero separately. They define the properties of the parallel Alfv\'{e}n (A-mode, $\Lambda_{11}=0$), parallel X-mode ($\Lambda_{22}=0$), and parallel longitudinal mode (L-mode, $\Lambda_{33}=0$). As only longitudinal waves are present in our simulations, we are interested in the solutions to the latter equation, $\Lambda_{33}=0$, with $\Lambda_{33}$ defined as: \citep{Rafat2019b,Rafat2019c}
\begin{equation}\label{eq-a4}
 \Lambda_{33} = K_{33} = 1 - \frac{\omega_\mathrm{p}^2}{\omega^2} z^2 W(z) = 0.
\end{equation}
In order to account for the relativistic pulsar conditions, a Maxwell-J\"{u}ttner distribution is suited to model the velocity distribution function for the background and beam plasma, respectively (Equations~\ref{eq2}-\ref{eq3}). These distribution functions, labeled $g_\alpha(u)$ ($\alpha = 0,1$ as subscripts for background and beam, respectively) define the respective relativistic plasma dispersion functions $W_\alpha(z)$:
\begin{equation}\label{eq-a5}
 W(z) = W_0(z) + r_\mathrm{n} \gamma_\mathrm{b} W_1(z),
\end{equation}
\begin{equation}\label{eq-a6}
 W_\alpha(z) = \frac{1}{n_\alpha} \int_{-\infty}^{\infty} \mathrm{d}u \frac{ \frac{\mathrm{d}g_\alpha(u)}{\mathrm{d}u} }{\beta - z}.
\end{equation}
Here, $z$ is given by $z = \omega / ck_\parallel$, and $n_\alpha$ is the density of the respective species.

The equations \ref{eq-a4}-\ref{eq-a6} can be solved with numerical methods: Firstly, the real and imaginary parts of the complex function given by Equation~\ref{eq-a4} are analyzed independently. For a given value of $k$, at which the a root of the dispersion relation is expected to be found, each part is sampled on a grid spanned by the real and imaginary parts of the frequency $\omega$. An intersection of both zero contour levels represents a solution to the dispersion relation. A multi-dimensional root finding function in \texttt{Python} subsequently uses this value as an initial guess in order to find a solution of higher accuracy. To obtain a solution for a subsequent value of $k$, a linear extrapolation is used in order to provide an initial guess that serves as an input for the same root finding function, such that a subsequent triple of solutions is obtained. An iterative continuation of this procedure finally yields the respective curves of the real and imaginary parts of the frequency as a function of $k$. Further details on this method, for example in terms of resolution, can be found in ref.~\citet{Manthei2021}.

\section{Stability of the simulation results} \label{AppendixB}

\begin{figure*}
    \centering
    \gridline{
              \fig{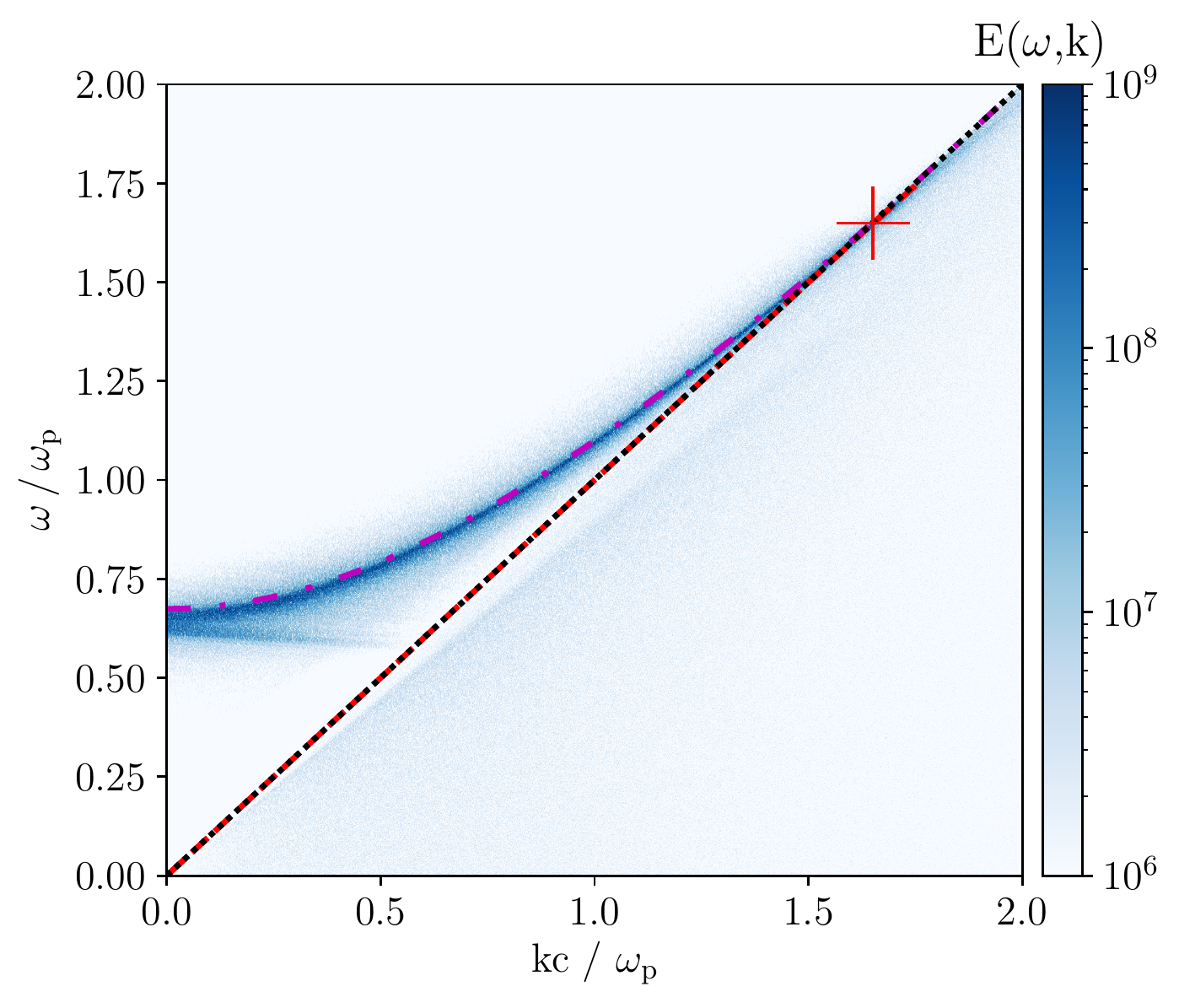}{0.49\textwidth}{(a) }
              \fig{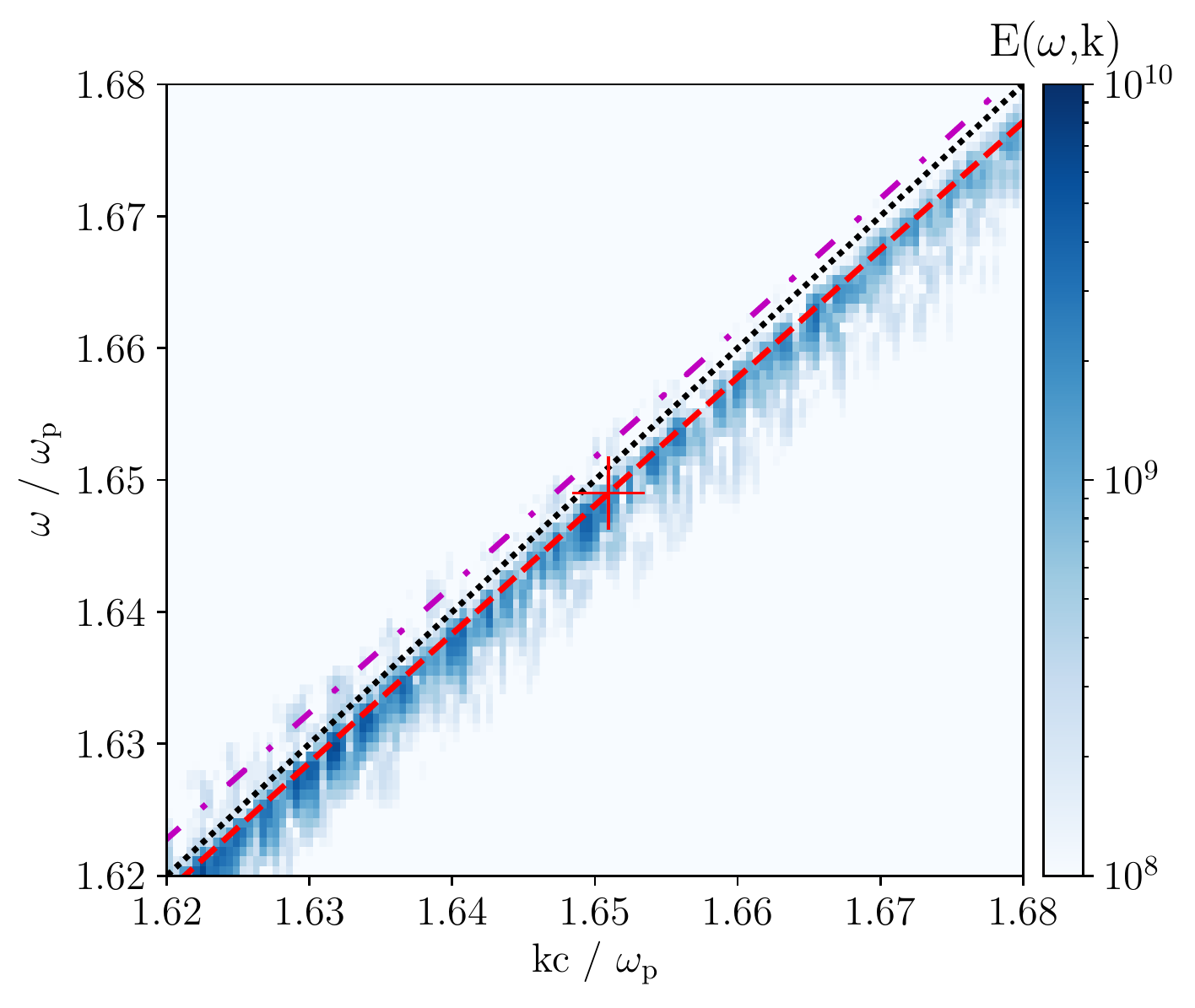}{0.49\textwidth}{(b) }
             }
    \caption{
    Dispersion diagram (a) and zoom-in region (b) for the simulation corresponding to the Case~2 and $20L$.
    The analytical linear dispersion modes (Equations~\ref{eq-a4}-\ref{eq-a6}) are overlaid.
    \textit{Black dotted line:} The light wave line, $\omega = c k$.
    \textit{Red dashed line:} The subluminal L-mode.
    \textit{Magenta dash-dotted line:} The superluminal L-mode.
    \textit{Red plus:} Analytically predicted position of the initially most unstable wave.
    }
    \label{fig-a1}
\end{figure*}

We especially tested the proper description of the formation of solitons in our simulations.
Specifically we investigated whether the simulated subluminal waves are sufficiently 
well separated from the dispersion line of the light waves in the Fourier space $\omega-k$.
For our tests, we selected Case~2 in which the subluminal waves are closest to the 
light-wave line and the resolution of the wavenumber resolution might have the largest 
impact. We increased the simulation domain length while the frequency resolution 
was sufficiently high in all cases.
Other parameters remained the same as described in Section~\ref{sec:methods}.
We ran simulations with $iL = (1, 2, 4, 10,20)\times L$,
where $L \approx 711\,d_\mathrm{e}$ is the original simulation length presented in the Results section.
The wave number resolution is
$\Delta k c / \omega_\mathrm{p} \approx 4.4\times 10^{-4}$
for $i=20$.
The estimated distance of the position (frequency and wavenumber) of the most unstable 
waves from the light wave line $\omega = c k$, predicted by the linear dispersion 
theory, is
$\Delta k_\mathrm{a} c / \omega_\mathrm{p} \approx 9\times 10^{-4} $
and
$\Delta \omega_\mathrm{a} / \omega_\mathrm{p} \approx 9 \times 10^{-4}$.

Figure~\ref{fig-a1} shows the dispersion diagram and a zoom-in region close to the 
position of the most unstable wave. 
The positions of the superluminal and subluminal modes calculated from the analytical Equations~\ref{eq-a4}-\ref{eq-a6} are overlaid.
The most unstable mode is well resolved and it is the subluminal mode.

\begin{figure}
    \centering
    \includegraphics[width=0.5\textwidth]{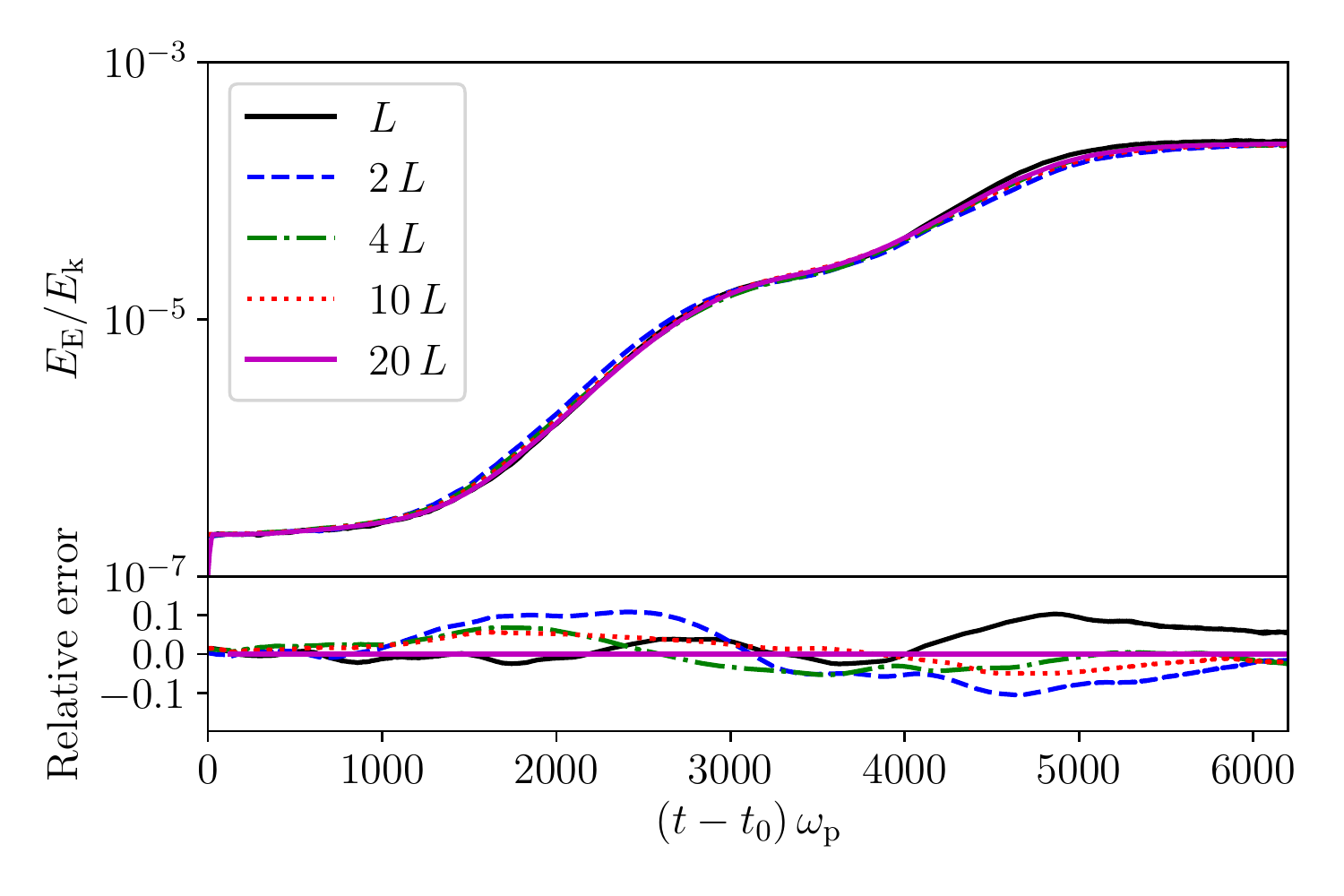}
    \caption{
    \textit{Top:} Evolution of the ratio between electrostatic and kinetic energy for simulation lengths $(1,2,4,10,20)\times L$.
    The increase of the simulation length allows to distinguish the subluminal waves from  the light wave line as the wavenumber resolution increases.
    \textit{Bottom:} Relative error estimated as the difference between the given simulation length and the largest simulation with length $20L$.
    }
    \label{fig-a2}
\end{figure}

Figure~\ref{fig-a2} presents the time evolution of the ratio of electrostatic and kinetic energy for 
different simulation lengths and their relative errors. 
The relative error is estimated as the relative difference between the simulation curve $E_\mathrm{E}/E_\mathrm{k}(iL)$, and the curve for simulation with the largest resolution $E_\mathrm{E}/E_\mathrm{k}(20L)$
\begin{equation}
  \mathrm{Relative~error~} (iL) = 
  \frac{ \frac{E_\mathrm{E}}{E_\mathrm{k}}\left(iL\right) - \frac{E_\mathrm{E}}{E_\mathrm{k}}(20L)}{\frac{E_\mathrm{E}}{E_\mathrm{k}}(20L)}, \quad i=1,2,4,10,20,
\end{equation}
with $iL$ the simulation length.
As the start time of the wave growth can slightly differ between simulation runs (not only for different simulation lengths, but also for runs with the same simulation setup if the particles are initialized with different random numbers),
the evolution profiles are slightly shifted in time by a factor $t_0$.
Although the simulation with size $L$ does not have a high enough wavenumber resolution to cover the difference between the light wave line and the subluminal mode,
the evolution and anticipated saturation energy of solitons,
which are formed almost at the end of the simulations,
are within the $\sim12\,\%$ error interval.



\bibliographystyle{aasjournal}
\bibliography{references}{}



\end{document}